\documentclass[10pt,onecolumn,showpacs,pre,preprintnumbers,amsmath,amssymb,aps,longbibliography]{revtex4-1}
\usepackage{lipsum,capt-of,graphicx,dcolumn,bm,color,comment,subfigure,upgreek,amsmath,amssymb}
\newcommand{\transp}{^{\rm T}}
\newcommand{\greekvektor}[1]{\mbox{\boldmath$#1$\unboldmath}}
\begin{document}
\preprint{Phys. Rev. Fluids}
\title{Growth mechanisms of perturbations in boundary layers over a compliant
  wall}
\author{M. Malik$^{1}$}
\author{Martin Skote$^{2}$}
\author{Roland Bouffanais$^{1}$} \email[Corresponding Author:
]{bouffanais@sutd.edu.sg}
\affiliation{$^1$Singapore University of Technology and Design, $^2$School of
  Mechanical and Aerospace Engineering, Nanyang Technological University,
  Singapore}
\date{\today}
\begin{abstract}
  The temporal modal and nonmodal growth of three-dimensional perturbations in
  the boundary layer flow over an infinite compliant flat wall is
  considered. Using a wall-normal velocity/wall-normal vorticity formalism,
  the dynamic boundary condition at the compliant wall admits a linear
  dependence on the eigenvalue parameter, as compared to a quadratic one in
  the canonical formulation of the problem.  As a consequence, the continuous
  spectrum is accurately obtained. This enables us to effectively filter the
  pseudospectra, which is a prerequisite to the transient growth analysis. An
  energy-budget analysis for the least-decaying hydroelastic
  (static-divergence, traveling-wave-flutter and near-stationary transitional)
  and Tollmien--Schlichting modes in the parameter space reveals the primary
  routes of energy flow. Moreover, the maximum transient growth rate increases
  more slowly with the Reynolds number than for the solid wall case. The
  slowdown is due to a complex dependence of the wall-boundary condition with
  the Reynolds number, which translates into a transition of the fluid-solid
  interaction from a two-way to a one-way coupling. Unlike the solid-wall
  case, viscosity plays a pivotal role in the transient growth. The initial
  and optimal perturbations are compared with the boundary layer flow over a
  solid wall; differences and similarities are discussed.
\end{abstract}
\maketitle


\section{Introduction}

%
The interaction between a compliant wall and fluid flow is of high interest
among researchers due to their relevance to drag-reduction problems and
biofluid
mechanics~\cite{carpenter2000hydrodynamics,grotberg2004biofluid}. Such
interest arose following the pioneering experiments
by~\cite{kramer1957boundary}, and subsequent studies
by~\cite{chung1985composite}. Compliant walls also help delay transitions
caused by Tollmien--Schlichting waves, which was predicted
theoretically~\cite{carpenter1985hydrodynamic,davies1997numerical}, and
confirmed experimentally~\cite{gaster1988dolphin}. This interaction between
compliant wall and flow was theoretically modeled as wall-admittance
by~\cite{landahl1962stability}, while subsequent workers treated it as a
two-way coupling, in which the wall is regarded as a membrane with a
plate-like behavior that responds through its velocity field to the
flow-induced forcing in a typical fashion of fluid-structure
interaction~\cite{kornecki1978aeroelastic,carpenter1985hydrodynamic}. The
compliant wall has also been modeled in a more complex way as the upper wall
of one or more viscoelastic layers~\cite{dixon1994optimization}.

While the fluid-based Tollmien--Schlichting modes are inhibited by the wall
compliance, the elastic nature of the wall gives rise to new modes of
instability, namely static-divergence modes and traveling wave flutter, which
are collectively known as \textit{hydroelastic
  modes}~\cite{gad1984interaction, carpenter1986hydrodynamic}. The phase-speed
of the traveling wave-flutter is found to be approximately coinciding with the
free wave-speed of the wall when neglecting both wall-damping and
fluid-forcing. The static divergence mode, has its origin in excessive
wall-damping.

Among the three types of modes, the Tollmien--Schlichting and the traveling
wave-flutter are prone to convective instability, and can be washed out
downstream by the mean flow. However, in the particular case of an infinite
compliant wall---when the natural phase-speed of the wall falls within the
range of the phase-speed of Tollmien--Schlichting waves, it is hard to
distinguish between instabilities due to hydroelastic modes and
Tollmien--Schlichting ones as they coalesce to form one single unstable mode
in the absolute sense~\cite{sen1988stability, gad1996compliant,
  davies2003convective} such that, if they are present at a certain location,
they remain there at all later time, while growing in amplitude and spreading
in space. Such different modes and their coalescence can also be identified in
a spatial stability calculation of a flow over finite compliant panels. Such
mode-coalescence phenomenon has been tracked in the wave-number plane, and its
absolutely unstable nature established~\cite{wiplier2001absolute}.  This
instability due to hydroelastic modes occur at low Reynolds numbers---much
smaller than the critical Reynolds number of an otherwise rigid wall, but with
a growth rate much smaller than the one of Tollmien--Schlichting waves. Due to
the presence of such hydroelastic modes, even at low Reynolds numbers,
compliant panels---with streamwise-length optimized between the growing
hydroelastic modes and inhibited Tollmien--Schlichting modes---are prescribed
for the purpose of
drag-reduction~\cite{carpenter1993optimization,dixon1994optimization}. Recently,~\cite{tsigklifis2015global}
has studied the global (temporal) traveling-wave-flutter modes in the presence
of finite-size compliant panels through a hybrid numerical technique where the
temporal eigensystem is provided with an input of spatial eigenvalues from a
separate spatial stability calculation. They have also studied the transient
growth of a superposition of normal modes.

In this paper, we perform for the first time a transient growth analysis of
the problem of hydrodynamic stability of a zero-pressure gradient boundary
layer flow over a flat plate with a normal-velocity/normal-vorticity formalism
for the study of three-dimensional (3D) modes. This formalism yields a
tractable linear-eigenvalue problem without the need to resort to the
traditional companion-matrix formulation, hence avoiding an unnecessary
increase in the number of unknowns (see
e.g.,~\cite{yeo1996absolute,wiplier2001absolute,hoepffner2010mechanisms,tsigklifis2015global}).
Our new alternative formalism results in the linear appearance of an otherwise
quadratic form of the eigenvalue parameter. By means of a specific spectral
method formulation for the continuous spectrum, we have access to more
accurate members of a superposition-state (i.e. a non-eigenstate) for the
first time, which is instrumental in performing the transient growth analysis
for this problem. We identify different types of modes and their associated
eigenfunctions, such as the static divergence, traveling-wave-flutter and
Tollmien--Schlichting modes, and present the unstable regions in the parameter
space.

A complete transient-growth study is conducted, which is supplemented with an
energy budget analysis of both modal and nonmodal growths. The nonmodal
evolution of a state of superposition of modes is analyzed in the plane of
wavenumbers. This allows us to identify, for the first time, the optimal
perturbation structure associated with maximum transient growth. This analysis
of nonmodal evolution is important as it leads to the identification of
optimal perturbation structures, which can be anticipated to be present in a
bypass transition. To this aim, we follow the traditional method developed
by~\cite{butler1992three} and \cite{trefethen1993hydrodynamic} (See also,
\cite{schmid2001stability}).

\section{Linear problem formulation and numerical approach}
\label{sec:linear}
\begin{figure}[htbp]
  \centering
  \includegraphics[width=.3\textwidth]{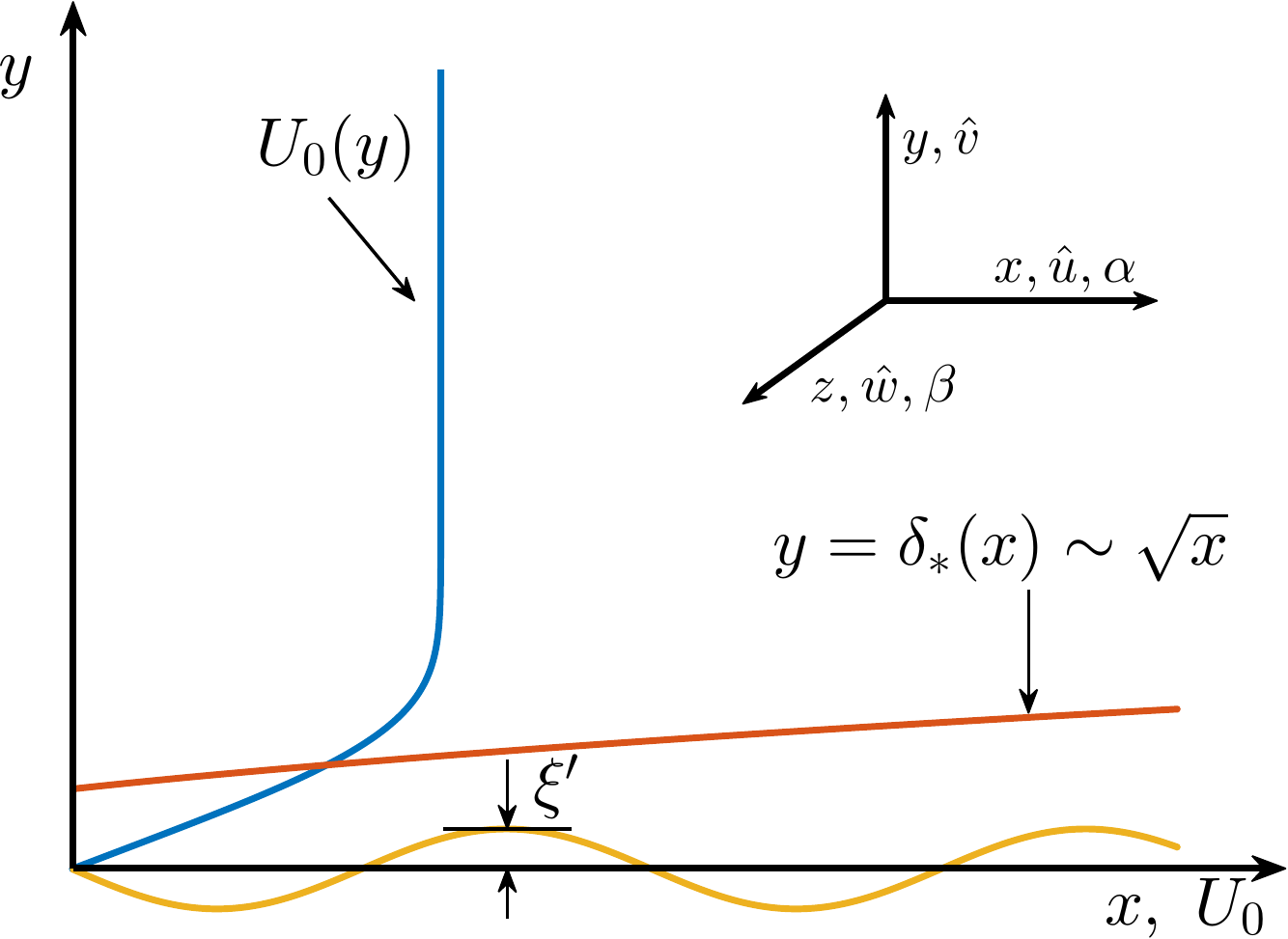}
  \caption{Schematic diagram: the thick solid line represents the
    mean-velocity profile with respect to $y$; dashed line shows a
    modal-perturbed compliant wall; dotted line is the displacement-thickness
    with respect to $x$.}
  \label{fig:sketch}
\end{figure}

Let $x$, $y$ and $z$ be the streamwise, normal and spanwise directions,
respectively, and let $(\hat{u}, \hat{v}, \hat{w})\transp$ be the velocity
fluctuations to the mean flow $(U_0(y), 0, 0)\transp$ in the respective
directions, where the mean flow is described under the approximation of
parallel flow with $U_0(y)$ obtained from the solution to the Blasius equation
(see Fig.~\ref{fig:sketch}). We study the flow stability using as variables
the fluctuations in normal-velocity, $\hat{v}$ and normal-vorticity
$\hat{\eta}\equiv \partial \hat{u}/\partial z - \partial \hat{w}/\partial
x$. This has the advantage of reducing the order of the discretized matrix of
the resulting eigensystem~\cite{schmid2001stability}.
Let $\mathbf{\hat{q}} = (\hat{v}, \hat{\eta})\transp$, and
$\mathbf{\hat{q}}=\mathbf{q'}(y)\exp [i(\alpha x + \beta z -\omega t)]$ where
$\alpha$, $\beta$ and $\omega$ are the streamwise and spanwise wavenumbers,
and complex frequency, respectively.  The amplitude $\mathbf{q'}$ is governed
by a system of Orr--Sommerfeld (OS) and Squire
equations~\cite{schmid2001stability}:
\begin{equation}
  \left (
    \begin{array}{cc}
      A_{11} & 0\\
      \beta DU_{0} & A_{22}
    \end{array}
  \right )
  \left(
    \begin{array}{c}
      v'\\
      \eta'
    \end{array}
  \right)
  =\omega
  \left (
    \begin{array}{cc}
      {\cal C} & 0\\
      0 & 1
    \end{array}
  \right )
  \left(
    \begin{array}{c}
      v'\\
      \eta'
    \end{array}
  \right),
  \label{ossq}
\end{equation}
where $A_{11} = \alpha (U_0 {\cal C} + D^2U_{0}) -i{\cal C}^2/Re$, $A_{22} =
\alpha U_0 -i{\cal C}/Re$, ${\cal C}=k^2 - D^2$, $D = d/dy$, and $k^2 =
\alpha^2 + \beta^2$. The Reynolds number reads as $Re = U_{\infty}
\delta_{*}/\nu$, where $U_\infty$, $\delta_{*}$ and $\nu$ are the freestream
velocity, displacement thickness and kinematic viscosity, respectively. The
$\omega$'s for this semi-bounded domain are either continuous or discrete. We
compute samples of the continuous spectrum via a different set of boundary
conditions for accuracy as will be discussed in latter sections. For the
continuous part of the spectrum, the complex frequencies are given by
\begin{equation}
  \omega = \alpha -i(k^2 + \tilde{k}^2)/Re,
  \label{cont_eigval}
\end{equation}
where $\tilde{k}\in \mathbb{R}^+$ represents the wavenumber in normal
direction of the continuous eigenfunctions in the
freestream. Equation~\eqref{cont_eigval} is obtained by relaxing the condition
that it is sufficient for $\mathbf{q'}$ to remain finite for $y \rightarrow
\infty$~\cite{grosch78:_orr}. Upon substituting $\omega$ from
Eq.~\eqref{cont_eigval} into Eq.~\eqref{ossq}, we get the eigensystem for the
continuous spectrum with $\tilde{k}^2$ as the eigenvalue.

\subsection{Boundary conditions at the compliant wall}
\label{sec:boundary}

%
The dynamics of the compliant wall, $y_d = \hat{\xi}_d(x_d,z_d)$ is given by
\begin{equation}
  m_d\frac{\partial^2\hat{\xi}_d}{\partial t_d^2} + d_d\frac{\partial
    \hat{\xi}_d}{\partial t_d} +
  \left[B_d \nabla_d^4 - T_d\nabla_d^2 + K_d\right]\hat{\xi}_d =
  -\hat{p}_d(y_d = 0) + 2\mu \left(\frac{\partial \hat{v_d}}{{\partial  y_d}}\right)_{y_d = 0},
  \label{membrane_eqn_dim}
\end{equation}
where $\nabla_d^2 = \partial_{x_dx_d}+\partial_{z_dz_d}$, and $m_d$, $d_d$,
$B_d$, $T_d$ and $K_d$ are the wall properties, namely, surface density,
damping coefficient, flexural rigidity, wall tension and stiffness constants,
respectively. The subscript ``{\scriptsize \it d}" is used to indicate that
quantities have their dimensions. The usual nondimensional equation used
(e.g., see~\cite{hoepffner2010mechanisms}) is
\begin{equation}
  m\hat{\xi}_{tt} + Re^{-1} d \hat{\xi}_{t} + Re^{-2}\left[B\nabla^4 - T\nabla^2 + K
  \right]\hat{\xi} = \hat{\sigma}_{yy},
  \label{membrane_eqn_unused}
\end{equation}
where $\hat{\sigma}_{yy} = -\hat{p}_{y=0} + 2Re^{-1}D\hat{v}|_{y = 0}$. These
equations arise from the following scalings: $m = m_d/(\rho_d \delta_{*})$, $d
= d_d\delta_{*}/(\nu \rho_d)$, $B = B_d/(\rho_d\delta_{*}\nu^2)$, $T =
T_d\delta_{*}/(\rho_d\nu^2)$, and $K = K_d\delta_{*}^3/(\rho_d\nu^2)$. Note
that the nondimensional values of all wall properties change with
$\delta_{*}$. In a parametric study, where variations in $Re$ are only due to
variations in $\delta_{*}$, such nondimensionalization will make $m$, $d$,
$B$, $T$ and $K$ vary with respect to $Re$. Therefore, following Yeo et
al.~\cite{yeo1996absolute}, we opt for a nondimensionalization of the wall
properties as $m = m_d/(\rho_d L_d)$, $d = d_d/(\rho_dU_{\infty})$, $B =
B_d/(\rho_dU_{\infty}^2L_d^3)$, $T = T_d/(\rho_dU_{\infty}^2L_d)$, $K =
K_dL_d/(\rho_dU_{\infty}^2)$, where $L_d = N_{\text{\scriptsize \it
    Re}}\nu/U_{\infty}$. Here, $N_{\text{\scriptsize \it Re}}$ is a number of
our choice used to fix the length-scale $L_d$ to a constant. Without any loss
of generality, we fix the length-scale $L_d$ by choosing $N_{\text{\scriptsize
    \it Re}} = 500$. Such a choice of the length-scale as a constant helps in
reformulating the boundary condition for Eq.~\eqref{membrane_eqn_unused},
thereby highlighting its explicit dependence on the Reynolds
number~\cite{yeo1996absolute}. Upon using these reference scales, and after
substituting the normal modes $(\hat{\xi}, \hat{\sigma}_{yy})^{\rm T} = (\xi',
\sigma'_{yy})^{\rm T}\exp [i(\alpha x + \beta z -\omega t)]$,
Eq.~\eqref{membrane_eqn_dim} reads as
\begin{equation}
  -m\gamma^{-1}\omega^2\xi' -id\omega \xi' + C(k,\gamma)\xi' = \sigma'_{yy},
  \label{membrane_eqn}
\end{equation}
where $C(k,\gamma) = Bk^4/\gamma^{3} + Tk^2/\gamma + K\gamma$ and $\gamma =
Re/N_{\text{\scriptsize \it Re}}$. It is worth noting that the scaling of $m$
by $\gamma$ signifies the fact that as one moves downstream, the inertia of
the wall becomes relatively less important because of the growth of the
boundary layer, which forces us to consider a larger volume of fluid, thus a
higher inertia of the fluid.

The kinematic conditions in the primed variables read as,
\begin{align}
  u'(0) &= -\xi'DU_{0}(0), \label{kinematic_1}\\
  v'(0) &= -i\omega \xi', \label{kinematic_2}\\
  w'(0) &= 0. \label{kinematic_3}
\end{align}
Note that Eq.~\eqref{kinematic_1} is obtained from a first-order Taylor
expansion of the no-slip condition, $u'(y = \xi') = 0$. Now, we are able to
obtain the wall surface pressure from the $x$- and $z$-momentum equations
evaluated at the wall:
\begin{align}
  -i\omega u'(0) + DU_{0}v'(0) &= -i\alpha p'(0) + Re^{-1}(D^2-k^2)u'|_{y = 0} \label{xmom_wall_prime}\\
  -i\omega w'(0) &=-i\beta p'(0) + Re^{-1}(D^2-k^2)w'|_{y =
    0}.\label{zmom_wall_prime}
\end{align}
The LHS of Eqs.~\eqref{xmom_wall_prime} and~\eqref{zmom_wall_prime} are zero
due to the kinematic conditions given by
Eqs.~\eqref{kinematic_1}--\eqref{kinematic_3}. Upon adding
Eq.~\eqref{xmom_wall_prime} and Eq.~\eqref{zmom_wall_prime}, the wall-pressure
is given by
\begin{equation}
  p'(0) = (Re k^2)^{-1}(D^2 - k^2)\left [ Dv' + r \eta' \right ]_{y = 0}, \label{p_wall}
\end{equation}
where $r = (\alpha - \beta)/(\alpha + \beta)$. Equation~\eqref{p_wall}
suggests that the perturbed wall-pressure vanishes in the inviscid limit. This
is due to the creeping nature of the flow for which there is a complete
balancing of pressure gradients in the streamwise and spanwise directions by
viscous forces. Hence, in such a limit, the fluid-solid interaction occurs via
a one-way coupling, i.e., in the linear first-order case the dynamics of the
wall affects the flow without itself being affected by the flow field. Note
that this has an important implication on the transient growth as will be
explained later.

The amplitudes $u'(y)$ and $w'(y)$ are obtained by continuity and by the
definition of $\eta'$ as $u' = ik^{-2}(\alpha Dv' -\beta \eta')$ and $w' =
ik^{-2}(\beta Dv' + \alpha \eta')$.  At the wall, using Eq.~\eqref{p_wall} for
$p'$, one can easily recast $\sigma'_{yy} = -p'(0) + 2Dv'(0)/Re$ as
\begin{equation}
  \sigma'_{yy} = Re^{-1}\left [\left (3D - k^{-2}D^3\right ) v' - rk^{-2} (D^2 - k^2)\eta'    \right ]_{y = 0}.
  \label{sigmaprime}
\end{equation}
Now, upon substituting these expressions of $\sigma'_{yy}$, $u'$, $w'$ and
$\xi'$ into Eqs.~\eqref{membrane_eqn}, \eqref{kinematic_1} and
\eqref{kinematic_3}, we derive the boundary conditions for $v'$ and $\eta'$ at
the wall as
\begin{align}
  -\frac{i\omega m k^2}{\gamma}v' & = \left[ -dk^2 + \frac{i\alpha C(k,
      \gamma)}{DU_{0}}D +\frac{3k^2}{Re}D-\frac{D^3}{Re} \right]v' - \left [
    \frac{i\beta C(k,\gamma)}{DU_{0}} + \frac{r}{Re} (D^2-k^2) \right ]\eta',
  \label{dynamic}\\
  \omega [-\alpha Dv' + \beta\eta'] &= k^2 DU_{0}v', \label{u_prime_wall_xiU0y}\\
  \beta Dv' + \alpha \eta' &= 0 \label{eqn_w_w_0}.
\end{align}
Note that in Eq.~\eqref{dynamic}, the eigenvalue $\omega$ appears linearly,
although it appeared as a quadratic term in the original
Eq~\eqref{membrane_eqn}. This is a direct consequence of the following three
substitutions in the LHS of Eq.~\eqref{membrane_eqn}: $\{\omega^2\xi', \omega
\xi, \xi' \} \leftarrow \{ i\omega v'(0), iv'(0), -u'(0)/DU_{0}(0)\}$. It is
worth highlighting that this step is crucial since it allows us to avoid a
nonlinear eigensystem, and thereby circumventing the need to resort to the
companion matrix method. This becomes possible as the linear system is
inherently not quadratic in the state variable of velocity field. This
formulation is clearly economical when it comes to obtaining the
eigenvalues. For instance, it drastically reduces the computational effort by
a factor of 8 when compared to the formulation
of~\cite{hoepffner2010mechanisms}, in which all four unknowns $\{u', v', w',
p'\}^{\rm T}$ are collocated across the entire flow domain---the number of
unknowns in the present case is only two~\cite{golub13:_matrix_comput}.  For
the case of the continuous spectrum, the wall-boundary conditions are obtained
by substituting $\omega$ from Eq.~\eqref{cont_eigval} into
Eqs.~\eqref{dynamic} and~\eqref{u_prime_wall_xiU0y}.

\subsection{Freestream condition for the discrete and continuous spectra}

%
In theory, the freestream boundary conditions are to be satisfied infinitely
away from the compliant wall. However, in practice a finite-size domain is
considered along with the approximate freestream boundary conditions, which is
dependent on the domain size.  When a very large domain is considered in the
normal direction, the flow at the upper boundary is subjected to the vanishing
boundary conditions, $v'(y_{\max}) = \eta'(y_{\max}) = 0$, where $y_{\max}$ is
defined by $y \in [0, y_{\max}]$. When the wavenumbers $\alpha$ and $\beta$
are very small, such vanishing boundary conditions satisfactorily apply only
for very large domains, e.g., for domains with $y_{\max} > 150$ in units of
displacement thickness. This is because, at such low wavenumbers, the weak
viscous effects require a long distance in the normal direction to damp out
the energy injected into the flow by the velocity field of the wall. We
verified numerically using a multi-point boundary value problem
solver~\footnote{Matlab \texttt{bvp5c}, which has the limitation that it can
  only give one eigenvalue and the corresponding eigenfunction.} that only for
such large domains with $y_{\max}\sim 150$, all higher derivatives of $v'$ and
$\eta'$ vanish for an eigenfunction from the discrete part of the
spectrum. However, from the numerical standpoint, when such large domains are
considered in combination with large numbers of collocation points or multiple
Chebyshev domains, the continuous part of the spectrum gets contaminated, and
yields a so-called pseudospectra, see
Refs.~\cite{schmid2001stability,trefethen2005spectra}.  It is therefore
imperative to find a way to enforce a set of freestream boundary conditions as
accurately as possible for domain sizes as small as $y_{\max} = 20$. To this
aim, we adopt two different sets of freestream boundary conditions for both
the discrete and continuous parts of the spectrum.

For the discrete part of the spectrum, the OS equation---i.e. the first
component of Eq.~\eqref{ossq}---admits the classical solution $v'(y) = \sum_{j
  =1}^4A_j e^{\lambda_j y},$ at the freestream where $\{A_j\}_{j=1,\cdots,4}$
are constants. The four $\lambda_j$'s are given by $\lambda_{1,3} = \mp k$ and
$\lambda_{2,4} = \mp \sqrt{i Re (\alpha - \omega) + k^2}$ as found
in~\cite{schmid2001stability}. Among these solutions, $\lambda_3$ and
$\lambda_4$ have a positive real-part leading to nonphysical and exponentially
growing modes in the limit of $y\rightarrow \infty$. These are trivially
eliminated by setting $A_3 = A_4 = 0$. The second mode is a decaying one, and
since $\lambda_2$ is proportional to $\sqrt{Re}$, it should pose no problem in
satisfying $v'(y_{\max}) = 0$ at relatively large Reynolds number.  However,
given $\lambda_1=-k$, it is possible that the first mode might not decay fast
enough to accurately satisfy this freestream condition. For instance, this
would be an issue when considering $y_{\max}=20$ and $k$ as small as
$0.05$---such a small value is not unrealistic at low $Re$, especially given
that $v'(y=0)\neq 0$ as in our case. Though these conditions may be satisfied
for large values of $y_{\max}$, a sufficiently satisfactory guess would lead
to a very large value for $y_{\max}$, which would result in prohibitively high
computational cost. To circumvent this critical issue, we propose a change to
the freestream boundary conditions so as to account for the specificities of
the exponential decay-rate of the first mode associated with $\lambda_1=-k$.

For the normal vorticity, it is rather difficult to prescribe the freestream
condition from its behavior $\eta'(y) = A_5 e^{\lambda_2 y} + \eta'_p(y)$ due
to the functional nature of $\eta'_p(y)$:
\begin{equation}
  \begin{split}
    \eta'_p(y) &=\frac{i\beta Re}{\lambda_4-\lambda_2}\left ( e^{\lambda_2y}
      \int_0^y DU_{0}(y') v'(y') e^{-\lambda_2y'} dy' - e^{\lambda_4y}
      \int_0^y DU_{0}(y') v'(y') e^{-\lambda_4y'} dy' \right
    ). \label{etap_fs}
  \end{split}
\end{equation}
However, one should note that $w'(0) = 0$, even in the presence of a compliant
wall. Therefore, this gives the motivation of retaining $w'(y_{\max}) = 0$ as
one normally does for the solid-wall case. Hence, the freestream conditions
for the discrete part of the spectrum are given by $ Dv'+kv' = 0, \ D^2v'+kDv'
= 0, \ \mbox{and} \ \beta Dv' + \alpha \eta' = 0 $ at $y=y_{\max}$.

In the case of the continuous spectrum, the freestream behavior is given by
\begin{align}
  v'(y) &= \tilde{A}_1\exp(i\tilde{k} y) + \tilde{A}_2\exp(-i\tilde{k} y) + \tilde{A}_3\exp(-ky), \label{vfreestreamcont}\\
  \eta'(y) &= \tilde{A}_4\exp(i\tilde{k} y) + \tilde{A}_5\exp(-i\tilde{k} y) +
  \eta'_p(y).
\end{align}
Reference~\cite{jacobs1998shear} says that the behavior given by
Eq.~\eqref{vfreestreamcont} can be written as $ (D^2 + \tilde{k}^2) v' =
\tilde{A}_3(k^2 + \tilde{k}^2) \exp(-ky), $ which they implemented as a
boundary condition, after removing $\tilde{A_3}$, and by evaluating the above
equation at two different locations in the freestream. Here, we prefer to
implement the above condition at a single location in the freestream upon
eliminating the constant $\tilde{A_3}$ from the derivatives. Consequently, the
freestream conditions for the continuous spectrum read as $ D^2(D+k)v' =
-\tilde{k}^2(D+k)v', \ D^3(D+k)v' = -\tilde{k}^2D(D+k)v', \ \mbox{and} \ \beta
Dv' + \alpha \eta' = 0.  $

\subsection{Numerical method}

%
We consider a domain such that $y \in [0, y_{\max}]$, with $y_{\max}=20$ in
the units of displacement thickness $\delta_*=\int_0^\infty (1-U_0(y))\,
dy$. We use a Chebyshev spectral method of order $N=300$ with $N+1$
Gauss--Lobatto collocation points to solve the eigensystems [Eqs.~\eqref{ossq}
and its continuous spectrum version]. To map the boundary layer into the
Chebyshev domain, we use $y = {a_1(1+y_c)}/{a_2-y_c}$ with $y_c \in [-1,1]$,
$a_1 = y_sy_b/(y_b-2y_s)$, $a_2 = 1 + 2a_1/y_b$. Note that $y_c$ can be any of
the Gauss--Lobatto points, namely $y_{c,j} = \cos\left([j-1]\pi/N\right), j =
1, \cdots, N+1$. The singularities originating from the boundary conditions
involving no $\omega$ nor $\tilde{k}^2$ were simply removed by means of
algebraic operations on the set of Chebyshev coefficients. To carry out the
integrations appearing in the following sections, we use the classical
quadrature rule $ \int_{0}^{\infty}f(y)dy = \sum_{j = 0}^N W_j f(y_j) \left (
  {dy}/{dy_c} \right )_j.  $ The weights $W_j$'s are given by $ W_j =
({b_j}/{N}) \{ 2+ \sum_{n=2}^N c_n [1+(-1)^{n}](1-n^2)^{-1}
\cos\left({nj\pi}/{N}\right) \}, $ with the coefficients
$\{b_j\}_{j=0,\cdots,N}$ given by $b_0 = b_N = 1/2$, and $b_j = 1 \ \mbox{for}
\ (1\leq j \leq N-1)$, and $\{c_n\}_{n=0,\cdots,N}$ given by $c_0 = c_N = 1$
and $c_n = 2 \ \mbox{for} \ (1\leq n \leq N-1)$~\cite{schmid2001stability}.

\section{Modal Analysis}
\label{sec:modal}

In the eigensystems for the discrete and continuous spectra, there is a total
of 8 parameters, namely $\{\alpha, \beta, Re, m, d, B, T, K\}$ characterizing
the flow and compliant wall. Arguably, an exhaustive study in such a large
parameter-space would be impractical, if not prohibitive. Therefore, we limit
our study by narrowing down to a specific region of interest in the
parameter-space. Specifically, we fix the wall parameters such that the flow
exhibits different patterns of unstable regions in the sub-parameter-space of
$(\alpha, \beta, Re)$.  The wall has its own phase-speed, $c_w$ which can be
calculated as:
\begin{equation}
  c_w = \Im \left({\sqrt{(\gamma d)^2 - 4m\gamma  C(k,\gamma)}}\right)/{(2m\alpha)},
  \label{Eqcw}
\end{equation}
which is, in fact, the imaginary part of the growth rate of the wall
perturbation (shown at a later stage of this paper in
Eq.~\eqref{wallgrowthrate} divided by the wavenumber). However, this
phase-speed is modulated by the damping force, which can be tuned by a
frequency dependent forcing term. Therefore, we consider the {\em free
  wave-speed} of the wall for fixing the wall parameters. The free wave-speed
of the wall, $c_w^{\text{\scriptsize free}}$ is given by
$c_w^{\text{\scriptsize free}} = \sqrt{\gamma C(k,\gamma)/m}/\alpha$, which is
obtained by setting $d \rightarrow 0$ in Eq.~\eqref{Eqcw}. If the absolute
value of $c_w^{\text{\scriptsize free}}$ is small---or within the range of an
order of magnitude of the flow speed---the wall can be termed as
\textit{soft}. Such soft walls can exhibit instabilities due to different
types of modes with more than one region of instability in the parameters
space as a consequence of the wall-flow interaction when the frequencies of
the wall and flow are in the same range. Hence, we fix the wall parameters to
that of a soft wall.
\subsection{Modal stability analysis}
\begin{figure}[htbp]
  \begin{center}
    \includegraphics[width=.9\textwidth]{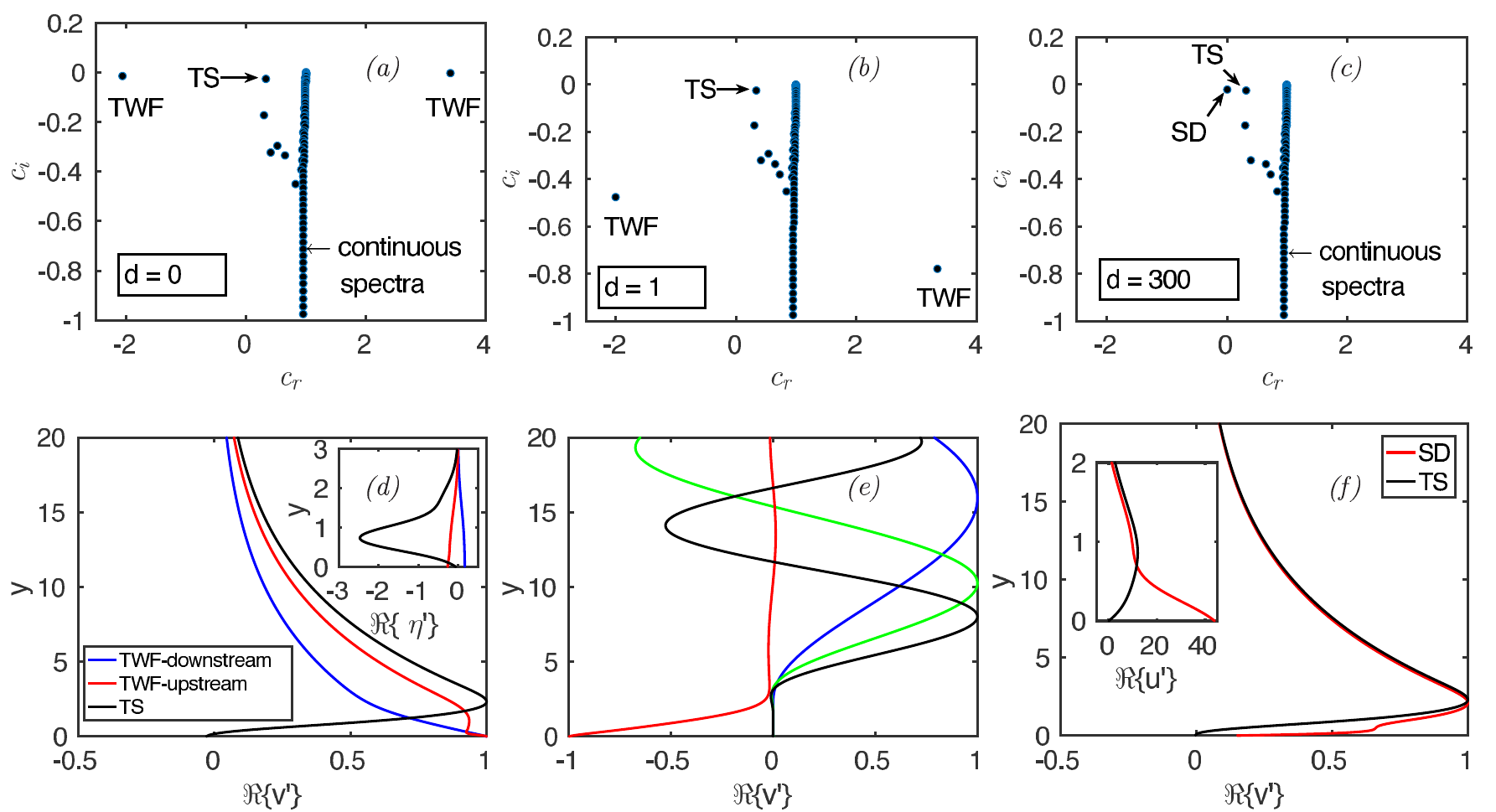}
    \caption{Spectra and eigenfunctions for $Re = 1000$, $\alpha = 0.1$,
      $\beta = 0.1$, $m = 2$, $K = 0.3$, $B = 3.2$, $T = 0.75$: (\textit{a})
      $d = 0$; (\textit{b}) $d = 1$; (\textit{c}) $d = 300$; (\textit{d})
      Normal velocity $v'$ and vorticity $\eta'$ of the marked discrete modes
      in panel (\textit{a}); (\textit{e}) Normal velocity of the
      least-decaying continuous modes in panel (\textit{a}); (\textit{f})
      Streamwise and normal components, $u'$ and $v'$ respectively, of the
      marked discrete modes in panel (\textit{c}).}
    \label{fig:spectra}
  \end{center}
\end{figure}

%
Figure~\ref{fig:spectra} shows the spectrum of the streamwise phase-speeds,
$c\equiv \omega/\alpha \ (\equiv c_r + i c_i)$. In this figure, we show three
types of characteristic modes, namely: (i) Tollmien--Schlichting (TS) modes,
(ii) Traveling Wave Flutter (TWF) modes, and (iii) Static-Divergence (SD)
modes that generally coexist in a flow over such compliant soft walls. Among
these modes, TWF and DS find their origin in the elastic and damping nature of
the wall, respectively.


%
Across the Figs.~\ref{fig:spectra}(\textit{a})--(\textit{c}), the
damping-coefficient increases. Figure~\ref{fig:spectra}(\textit{a}) shows two
TWF modes---one propagating upstream and the other one downstream---and the TS
mode that is the least decaying among all. The adopted method yields the
continuous spectra with extremely reduced distortion compared to previous
studies (e.g., see Fig. 3.4(\textit{a}) in~\cite{schmid2001stability}). As the
damping coefficient $d$ is increased from 0 to 1, one can note that the TWFs
decay faster (see Fig.~\ref{fig:spectra}(\textit{b})), and completely vanish
as $d$ is increased further (see Fig.~\ref{fig:spectra}(\textit{c})). At very
large $d$ values, there appears a new mode, the so-called
\textit{Static-Divergence} (SD) mode. We identify this mode as SD following
the experimental observation that such mode will have a phase-speed close to
zero, and that appears in the case of highly damped
walls~\cite{carpenter1986hydrodynamic}. Indeed, when $d$ is large, the inertia
is primarily balanced by the damping term. Therefore, Eq.~\eqref{dynamic}
gives rise to $\omega \approx 0$. This explains the origin of SD with a
phase-speed that is almost zero.

We note that when $\beta$ is gradually increased from zero, the phase-speed of
TWF modes increase as well (not shown here). This is tied to several facts:
(1) the TWF modes find their source at the wall, whose phase-speed $c_w$ is
given by Eq.~\eqref{Eqcw}, (2) these modes originate from the homogeneous part
of Eq.~\eqref{membrane_eqn}, and (3) there is a monotonously increasing
dependence of $c_w$ on $\beta$.

The eigenfunctions of a few selected sample modes are shown in
Fig.~\ref{fig:spectra}(\textit{d}-\textit{f}). Figure~\ref{fig:spectra}(\textit{d})
shows the velocity and vorticity components of the eigenfunctions of two TWF
modes and the TS mode of the spectrum shown in
Fig.~\ref{fig:spectra}(\textit{a}). The two TWF modes have normal vorticities
which are counter-rotating within the boundary layer. The downstream
propagating TWF (denoted as ``TWF-d") has a fast and monotonously decaying
eigenfunction with respect to the wall-normal coordinate $y$. Though such
monotonicity is absent in the case of the upstream propagating TWF mode
(denoted as ``TWF-u"), it exhibits maximum at the wall. This indicates that
the TWF modes are primarily inviscid in nature, and thereby approximately a
solutions to the Rayleigh equation
\begin{equation}
  (D^2 - \alpha^2)v'(y) = {D^2U_{0}}[U_0(y) - c]^{-1}v'(y).
  \label{RayleighEqn}
\end{equation}
The solution that decays exponentially is given
by~\cite{drazin1962instability}. The qualitative difference between the
profiles of $v'(y)$ for the TWF-u and TWF-d arises from the change in the sign
of the term $(U_0(y) - c)$ between these two modes. As the term $D^2U_{0}$ is
negative in the flow domain, the RHS of Eq.~\eqref{RayleighEqn} enhances the
decay of the TWF-d mode within the boundary layer. Intuitively, this effect
can be seen as amplifying the effect of the $-\alpha^2v'(y)$ on the LHS of
Eq.~\eqref{RayleighEqn}. In the case of the TWF-u mode, since $\alpha^2$ is
small, Eq.~\eqref{RayleighEqn} is equivalent to $[D^2 + f(y)]v'(y) = 0$ within
the boundary layer, where $f(y) > 0$, which precludes a decaying
solution. This phenomenon can be observed in the component $v'(y)$ of these
modes as is shown in Fig.~\ref{fig:spectra}(\textit{d}). This behavior of the
solutions for both the TWF-d and TWF-u modes can also be observed in a channel
flow, where one of the wall is a compliant (e.g., see Figs. 3(\textit{a}) and
3(\textit{b}) of~\cite{stewart2010local}), and can be explained through a
similar and analogous argument.

Figure~\ref{fig:spectra}(\textit{e}) shows a few eigenfunctions of the
continuous spectrum from Fig.~\ref{fig:spectra}(\textit{a}). In the case of
flat rigid wall, the freestream fluctuations of the continuous spectrum are
kept away due to a rapid quenching near the edge of the boundary layer, a
phenomenon known as {\it shear sheltering}~\cite{jacobs1998shear}. However, in
the present case, these fluctuations are allowed to penetrate the boundary
layer, but without the required sinusoidal oscillations in the normal
direction so as to match the nontrivial wall-boundary conditions.

Figure~\ref{fig:spectra}(\textit{f}) shows $v'(y)$ and $u'(y)$ for the TS and
SD modes shown in Fig.~\ref{fig:spectra}(\textit{c}). Though their normal
velocity component $v'(y)$ hardly differ, the streamwise components $u'(y)$ of
TS and SD modes differ notably. The SD mode exhibits a stronger streamwise
component $u'(y)$ near the wall.

\begin{figure}[htbp]
  \begin{center}
    \includegraphics[width=0.9\textwidth]{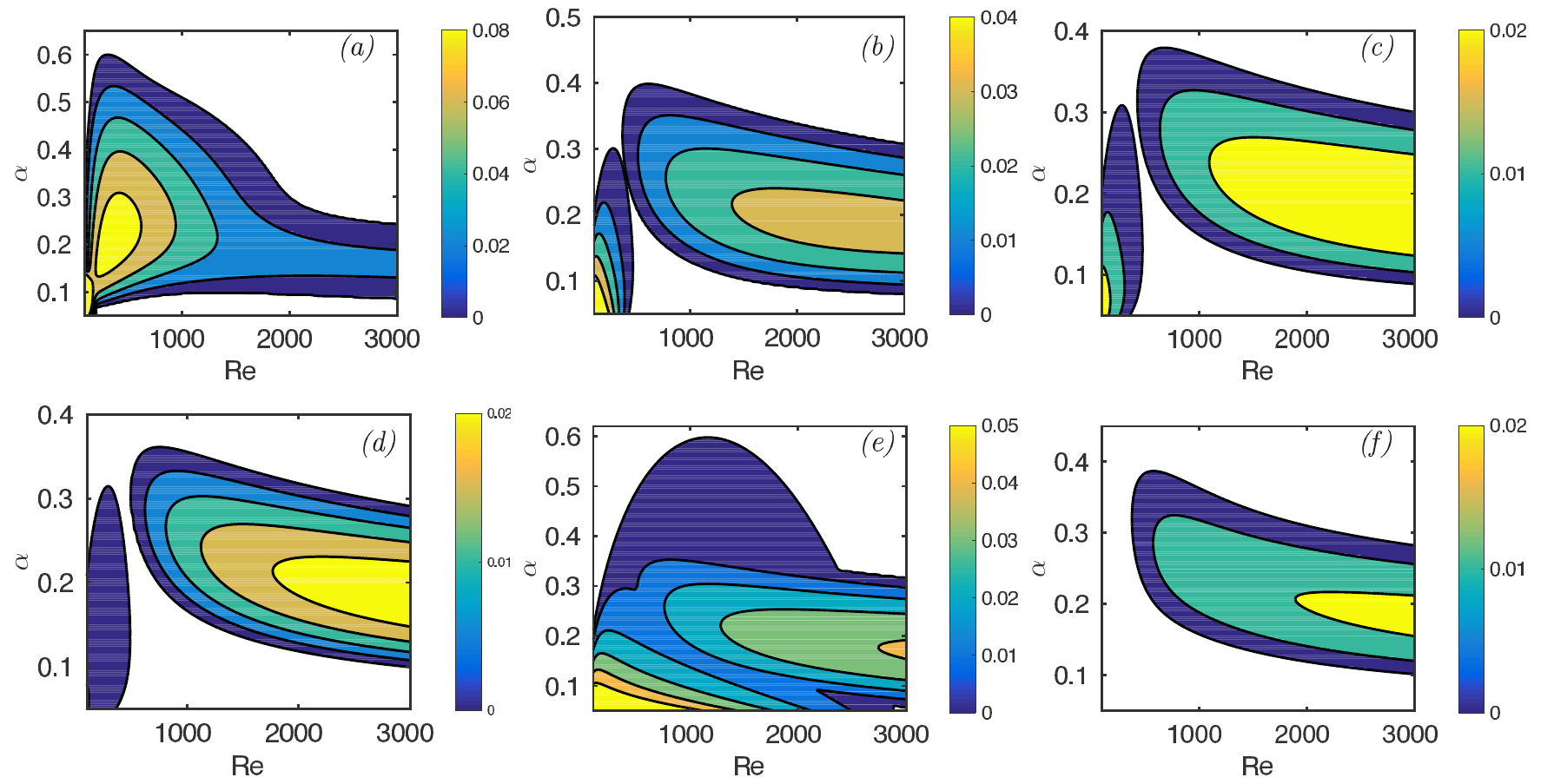}
    \vspace{-3ex}
    \caption{Contours of $c_i=\Im(c)$, with $c\equiv \omega/\alpha$, in the
      $Re-\alpha$ plane with $\beta = 0$, $T = 0.1$, $B = 0.2$ and $m=2$:
      (\textit{a}) $d=1$, $K = 0.05$; (\textit{b}), (\textit{c}) and
      (\textit{d}) are for $d = 10, 20, 100$ respectively with other
      parameters identical to (\textit{a}); (\textit{e}) $d=10$, $K = 0.005$,
      $B=0.2$, $T=0.1$; (\textit{f}) $d=10$, $K = 0.5$, $B=0.2$, $T=0.1$.}
    \label{fig:ci_rexal_comp}
  \end{center}
\end{figure}

The unstable regions are now presented in the parameter-spaces of $Re-\alpha$
and $\alpha-\beta$, for a chosen set of other parameters.
Figure~\ref{fig:ci_rexal_comp} shows the same contours but now for the flow
over a compliant wall with varying wall properties. In these figures, the
neutral curve (i.e., contour level $0.00$) serves as the separating boundary
between the unstable and stable regions.

\begin{figure*}[htbp]
  \begin{minipage}[t]{4.4in}
    \includegraphics[width=\textwidth]{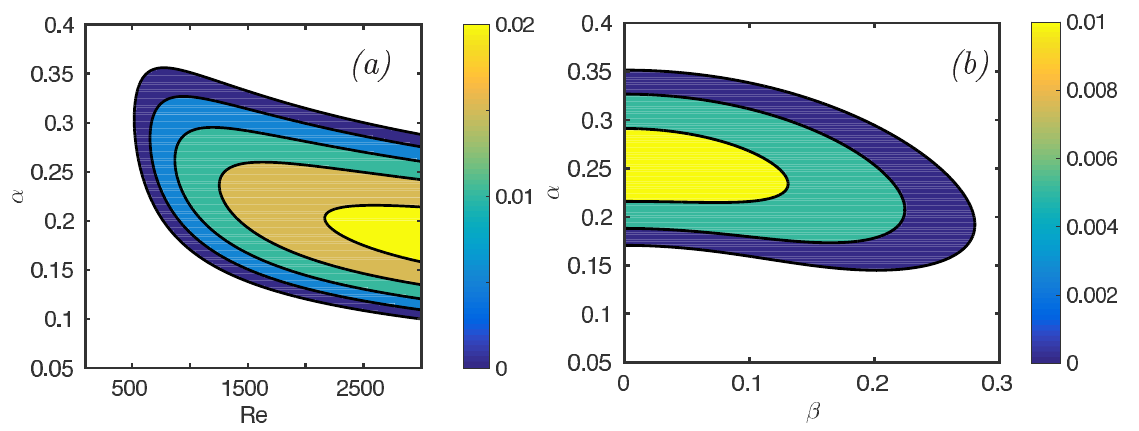}
  \end{minipage}
  \hspace{0.3in}
  \begin{minipage}[t]{2.2in}
    \includegraphics[width=\textwidth]{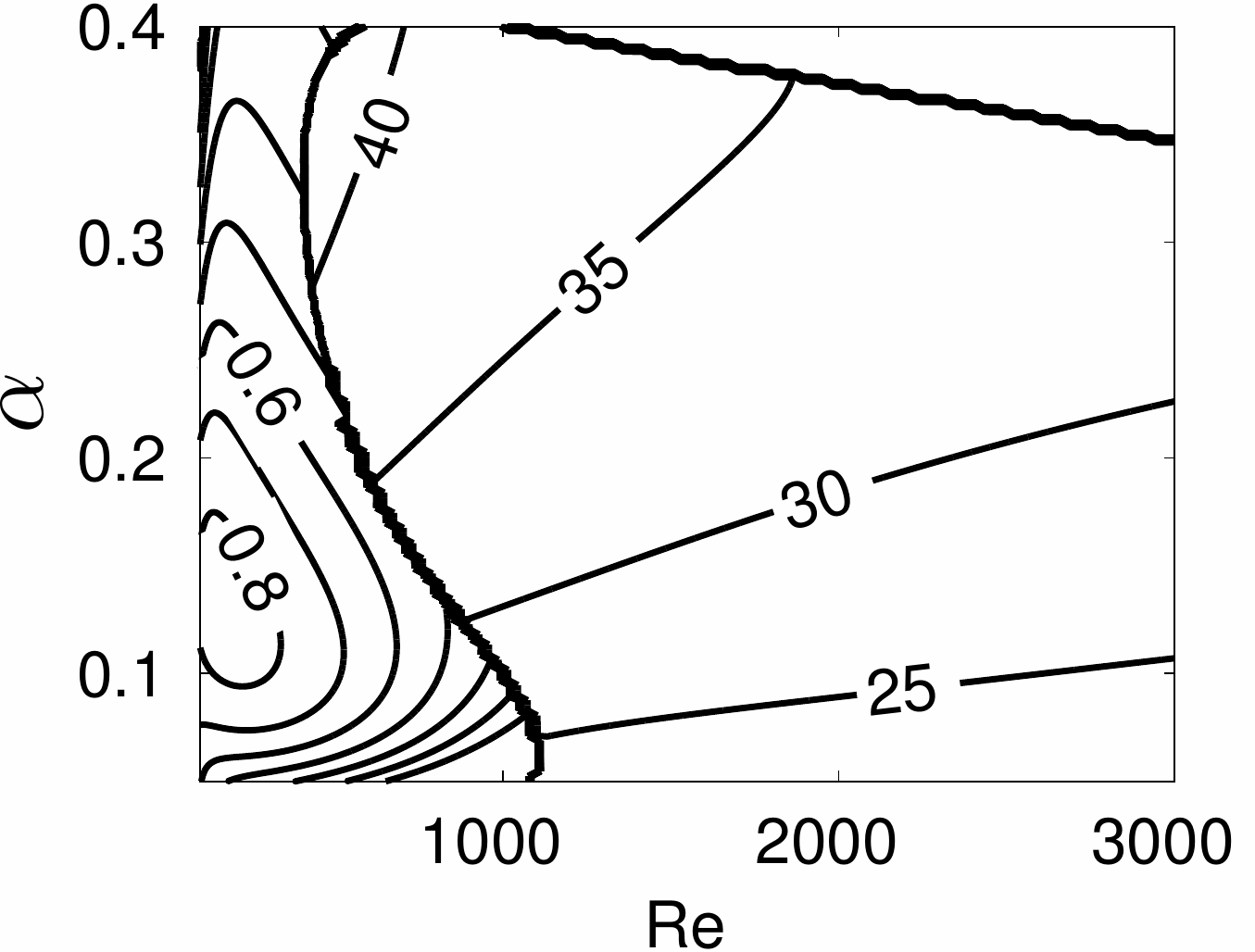}
  \end{minipage}
  \begin{minipage}[t]{4.3in}
    \vspace{-4ex}
    \captionof{figure}{Contours of $c_i=\Im(c)$ for a rigid wall: (\textit{a})
      in $(Re,\alpha)$ plane with $\beta = 0$; (\textit{b}) in
      $(\alpha,\beta)$ plane with $Re = 1000$.}
    \label{fig:c_i_rigid}
  \end{minipage}
  \begin{minipage}[t]{0.4in}
    \hspace{0.4in}
  \end{minipage}
  \begin{minipage}[t]{2.2in}
    \vspace{-4ex}
    \captionof{figure}{Contours of $c_r \times 100$ for the case of
      Fig.~\ref{fig:ci_rexal_comp}(\textit{c}).}
    \label{fig:cr_rexal_comp}
  \end{minipage}
\end{figure*}

Across
Figs.~\ref{fig:ci_rexal_comp}(\textit{a})--\ref{fig:ci_rexal_comp}(\textit{d}),
the damping coefficient $d$ increases while the other wall parameters have
been set to that of a soft wall except for
Fig.~\ref{fig:ci_rexal_comp}(\textit{f}). One may note that the flow can
develop instabilities for practically all values of the Reynolds number,
except for Fig.~\ref{fig:ci_rexal_comp}(\textit{f}) where the flow exhibits
stability at low $Re$. This is because
Figs.~\ref{fig:ci_rexal_comp}(\textit{a})--(\textit{e}) correspond to a soft
wall, while Fig.~\ref{fig:ci_rexal_comp}(\textit{f}) is obtained for a nonsoft
wall. The instability regions at low $Re$ in
Figs.~\ref{fig:ci_rexal_comp}(\textit{a})--(\textit{e}) arises from the
hydroelastic nature---e.g., due to the presence of unstable static divergence
modes when the damping parameter $d$ is large, or due to an unstable
near-stationary {\it transitional mode}, which appears due to the merging of
TS and TWF modes~\cite{sen1988stability,carpenter1990status} when $d$ is
small. In the higher $Re$ regime, the unstable modes are of the usual TS kind
and exists in the solid-wall case where these modes travel with a finite
phase-speed lower than the freestream velocity. This identification can be
justified based on the phase speeds of these modes which, will be presented
subsequently.

For low values of $d$, the unstable zone extends to a region including greater
values of $\alpha$. For the lowest value of $d$ considered ($d=1$ in
Fig.~\ref{fig:ci_rexal_comp}(\textit{a})), the regions of instability due to
the near-stationary transitional modes and the traveling TS modes merge to
form a larger unstable region in this parameter space. However, the dominant
unstable mode in this region is of the traveling TS kind. However, for larger
values of $d$, the instability regions of these two types of unstable modes
become distinct, where the near-stationary transitional unstable mode is
dominant at low $Re$, while the traveling TS mode dominates in the higher $Re$
regime as shown by
Fig.~\ref{fig:ci_rexal_comp}(\textit{d}). Figure~\ref{fig:ci_rexal_comp}(\textit{e})
is for a very low value of $K$ in comparison with
Figs.~\ref{fig:ci_rexal_comp}(\textit{a})--(\textit{d}). At such low value of
$K$, the phenomenon of larger region of the parameter space becoming unstable
is common with the situation for the case of low values for $d$ as shown in
Fig.~\ref{fig:ci_rexal_comp}(\textit{a}). However, there is a subtle
difference in the pattern of the values of the growth rate between
Figs.~\ref{fig:ci_rexal_comp}(\textit{a}) and~(\textit{e}). This difference
stems from the fact that decreasing $d$ increases the natural phase-speed of
the wall (See, Eq.~\ref{Eqcw}), while decreasing $K$ does the opposite. These
opposite natures of the wall-phase-speed are also reflected in the fluid-wall
coupled phase-speed ($c_r$) of the modes corresponding to
Fig.~\ref{fig:ci_rexal_comp}(\textit{a}) and~(\textit{e}) (not shown here). In
fact, the pattern in Fig.~\ref{fig:ci_rexal_comp}(\textit{e}) is similar to
the case of Fig.~\ref{fig:ci_rexal_comp}(\textit{b}), where these two differ
only in the values of $K$. Upon reducing $K$, the instability due to
near-stationary transitional modes merges with the region of instability
caused by the traveling TS mode, but differs from the case of
Fig.~\ref{fig:ci_rexal_comp}(\textit{a}) by not affecting the nature of
traveling TS mode being dominant in its corresponding location in the plane of
$Re$--$\alpha$.

Figure~\ref{fig:ci_rexal_comp}(\textit{f}) is for a high value of $K$. The
instability associated with the near-stationary transitional modes vanishes
altogether due to such high stiffness---the wall behaves more like a
flat-rigid one. Nonetheless, the critical Reynolds number of the viscous
instability is much lower in comparison with the rigid-wall case---the $c_i$
levels are significantly higher than those for the rigid-wall case at a chosen
$Re$. Furthermore, we observe that increasing $K$, $B$ and $T$ results in
recovering the results for the rigid-wall case since these parameters all tend
to increase the wall-phase-speed. However, it should be noted that among these
wall parameters, only $K$ has more pronounced effect, followed by $T$ and then
by $B$.

Upon comparison of Fig.~\ref{fig:ci_rexal_comp}(\textit{f}) with the
corresponding contours for a non-compliant rigid wall shown in
Fig.~\ref{fig:c_i_rigid}(\textit{a}), it should be noted that the flow over
the compliant wall, in the rigid-wall limit, exhibits higher instability than
the corresponding flow in the non-compliant rigid wall case. The instabilities
in both figures are due to TS modes. The difference in growth rates can be
explained through the classification due to~\cite{benjamin1963threefold}
and~\cite{landahl1962stability}. Under this classification, TS and TWF modes
belong to Class A and Class B, respectively, while the near-stationary
transitional and SD modes fall into the Class C category. Therefore, any
phenomenon that stabilizes Class-A modes will destabilize Class-B ones, and
\textit{vice versa}~\cite{carpenter1990status}. Since the wall damping in
Fig.~\ref{fig:ci_rexal_comp}(\textit{f}) stabilizes the TWF mode (see
Fig.~\ref{fig:spectra}), it exhibits enhanced TS mode instability when
compared to the non-compliant rigid-wall case.

\begin{figure*}[htbp]
  \begin{minipage}[t]{2.3in}
    \includegraphics[width=0.9\textwidth]{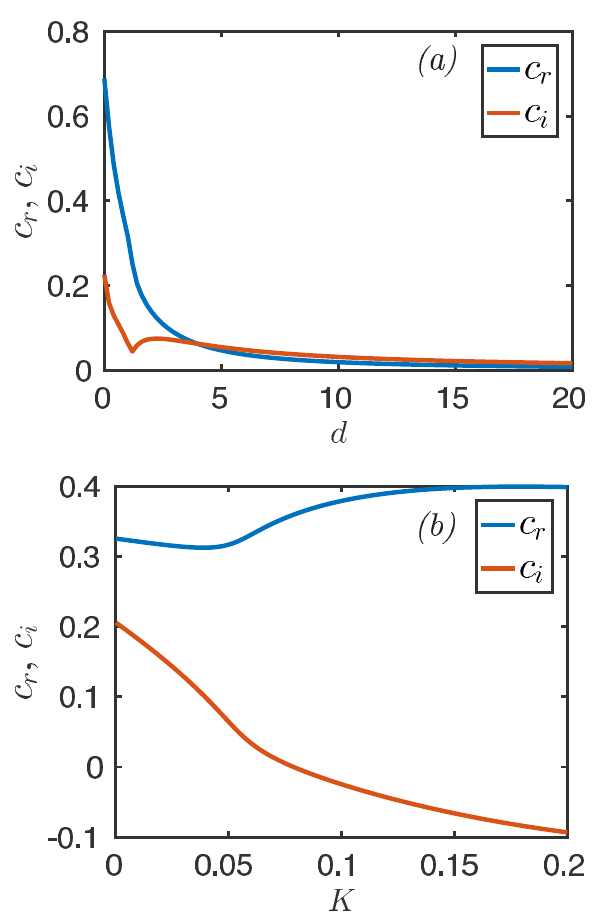}
  \end{minipage}
  \begin{minipage}[t]{4.6in}
    \includegraphics[width=0.9\textwidth]{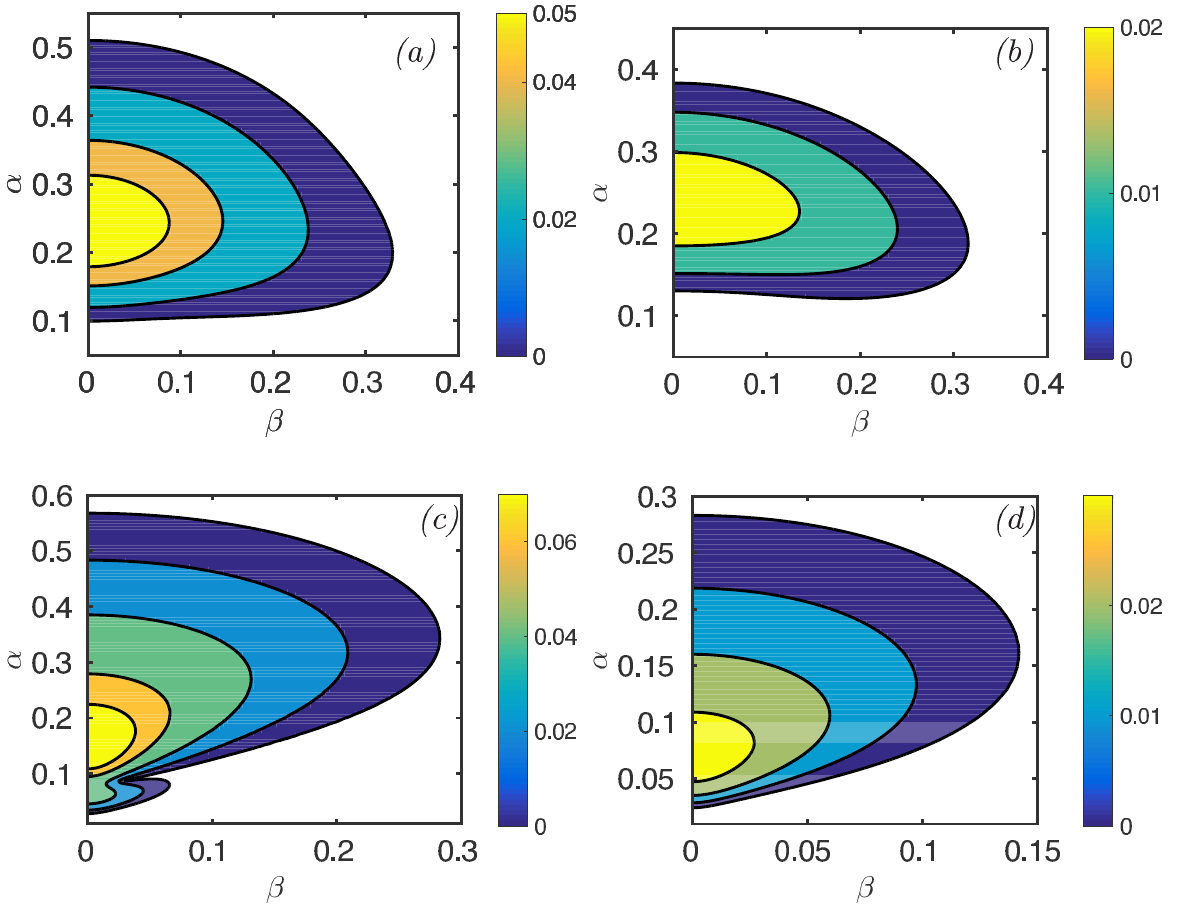}
  \end{minipage}
  \begin{minipage}[t]{2.4in}
    \vspace{-3ex}
    \captionof{figure}{$c_r$ and $c_i$ for $Re = 200$, $\alpha = 0.1$, $\beta
      = 0$, $m = 2$, $B = 0.2$, $T = 0.1$: (\textit{a}) versus $d$ with $K =
      0.05$; (\textit{b}) versus $K$ with $d = 1$.}
    \label{ev_d0d20}
  \end{minipage}
  \begin{minipage}[t]{0.2in}
    \hspace{0.2in}
  \end{minipage}
  \begin{minipage}[t]{4.3in}
    \vspace{-3ex}
    \captionof{figure}{Iso-contours of $c_i$ in $\alpha-\beta$ plane with $K =
      0.05$, $B =0.2$, $T= 0.1$: (\textit{a}) $Re = 1000$ and $d = 1$;
      (\textit{b}) $Re =1000$ and $d = 10$; (\textit{c}) $Re = 200$ and $d =
      1$; (\textit{d}) $Re= 200$ and $d = 10$.}
    \label{fig:ci_alxbt_comp}
  \end{minipage}
\end{figure*}

Figure~\ref{fig:cr_rexal_comp} shows the streamwise phase-speeds ($\Re\{c\}$)
corresponding to Figs.~\ref{fig:ci_rexal_comp}(\textit{c}). One may observe
that the unstable modes at low $Re$ are almost stationary, which is further
verified for other set of parameters in Fig.~\ref{fig:ci_rexal_comp}. However,
their phase-speed increases with decreasing $d$ as can be seen in
Fig~\ref{ev_d0d20}(\textit{a}), which shows the unstable mode for a range of
the parameter $d$ varying from $0$ to $20$. As $d$ is increased, the traveling
TS mode gets converted into a near-stationary transitional
mode. Figure~\ref{ev_d0d20}(\textit{b}) shows the evolution of the phase-speed
and growth-rate of a TS mode with the stiffness constant $K$. In this figure,
the least-decaying (or equivalently the fastest-growing) mode shown is not
stationary---i.e., the phase-speed is not close to zero---due to the low value
of the damping coefficient $d$. Except for $K$, the rest of the parameters has
been set to that of Fig.~\ref{fig:ci_rexal_comp}(\textit{a}). As $K$
increases, the phase-speed saturates to a constant value, and the mode gets
stabilized gradually, thereby suggesting the destabilizing nature of a soft
compliant wall.

Figure~\ref{fig:ci_alxbt_comp} shows the unstable region in the $\alpha-\beta$
plane with other parameters set to that of a soft wall to ensure that we are
in the presence of near-stationary transitional unstable modes whenever $Re$
is low. Figures~\ref{fig:ci_alxbt_comp}(\textit{a})
\&~\ref{fig:ci_alxbt_comp}(\textit{b}) show that the instability due to the
traveling TS modes spreads to higher values of $\beta$ in comparison to the
flat-rigid-wall case (shown in Fig.~\ref{fig:c_i_rigid}(\textit{b}))
suggesting that the oblique modes are prone to more instability due to wall
compliance. As observed in Fig.~\ref{fig:ci_rexal_comp}, the unstable
$\alpha$-band diminishes upon increasing $d$ as evident from the comparison of
Fig.~\ref{fig:ci_alxbt_comp}(\textit{a})
with~Fig.~\ref{fig:ci_alxbt_comp}(\textit{b}). The fact that these unstable
modes are of the traveling TS kind can also be verified from the contours of
phase-speeds (not shown here), which were all found to be above the level
$0.3$. The unstable region associated with the near-stationary modes at low
$Re$ has been shown in the $\alpha-\beta$ plane in
Figs~\ref{fig:ci_alxbt_comp}(\textit{c})
\&~\ref{fig:ci_alxbt_comp}(\textit{d}) at low values of $\alpha$ and
$\beta$. Although they appear only in the compliant-wall case, they exhibit a
clear signature of interaction with the flow through the shapes of these
contours. Indeed, if these instabilities are to be attributed prominently to
the wall, one would expect a circular symmetry of these contours since the
governing Eq.~\eqref{membrane_eqn} possesses such symmetry in the
$\alpha-\beta$ plane. One should note that for these modes at low $Re$,
$\alpha$ and $\beta$ can be understood from the variations of the phase-speed
of the wall shown in Eq.~\eqref{Eqcw}. Specifically, when both $\alpha$ and
$\beta$ are small---i.e., when $k$ is small, the contribution due to flexural
rigidity and tension are negligible. In such situations, Eq.~\eqref{Eqcw}
indicates that $c_w$ decreases with decreasing $Re$ or decreasing $K$. In
other words, for low values of $\alpha$ and $\beta$, the effect of a decrease
in $Re$ on the instability due to hydroelastic modes is similar to the effect
of a decrease in $K$. As can be seen from Fig.~\ref{ev_d0d20}(\textit{b}), a
decrease in $K$ causes instability, so is when $Re$ is decreased. This leads
to the observed unstable regime at low $Re$.

The comparison of Fig.~\ref{fig:ci_alxbt_comp}(\textit{c}) and
Fig.~\ref{fig:ci_alxbt_comp}(\textit{d}) implies that such destabilizing
interaction between the flow and the wall decreases with an increase of $d$ as
one would expect, it reduces the effective forcing by the fluid.

\subsection{Modal energy analysis}

%
We analyze the least-decaying mode in order to identify the root causes of the
instability mechanisms in terms of its constituent energies channeled through
different routes: e.g., via the meanflow, viscous dissipation, wall damping,
or the work done on the surface by the fluid. The total perturbation energy is
defined as
\begin{equation}
  E'(t) =\frac{e^{2\omega_it}}{2k^2} \left ( \int_{0}^{\infty} (k^2|v'|^2+ |Dv'|^2 + |\eta'|^2)dy + k^2\left ( {m}{\gamma}^{-1}|\omega \xi'|^2 + C(k,\gamma) |\xi'|^2 \right ) \right ).
\end{equation}
Given our formalism based on the set of working variables $(v',\eta')$, one
can rewrite this energy as
\begin{equation}
  E'(t) = \frac{e^{2\omega_it}}{2k^2}\int_{0}^{\infty}
  \left (
    \begin{array}{c}
      v' \\
      \eta'
    \end{array}
  \right )^{\transp}
  \left ( 
    \begin{array}{cc}
      k^2+D^{\rm T}D + \delta(y)\left [k^2\left (\frac{m}{\gamma}+\frac{C(k,\gamma)}{|\omega|^2} \right ) \right ] & 0\\
      0 & 1
    \end{array}
  \right )
  \left ( 
    \begin{array}{c}
      v'\\ \eta'
    \end{array}
  \right ) dy,
  \label{E_def}
\end{equation}
where $\delta(y)$ is the Dirac $\delta$-distribution. Using the identity $
\int_0^\infty|Dv'|^2dy = (v'^*Dv')_{y = 0} - \int_0^\infty v'^*D^2v' dy, $ and
Eq.~\eqref{ossq}, the rate-of-change of the perturbation energy reads as
\begin{equation}
  \frac{d E'}{d t} = \frac{d E'_{\text{\tiny W}}}{d t} - \frac{e^{2\omega_it}}{2k^2}\left (i\int_0^\infty\mathbf{q'}^{\transp}A\mathbf{q'} dy +i\omega v'^*(0)Dv'(0) + \text{c.c.}\right ),
  \label{E_teqn}
\end{equation}
where $\mathbf{q'} = (v', \eta')^{\rm T}$, $A$ is the coefficient-matrix on
the LHS of Eq.~\eqref{ossq}, ``c.c'' denotes the complex conjugate terms, the
superscript $^*$ corresponds to the complex conjugation operator, and $
E_{\text{\tiny
    W}}'=v'(0)^*\left({m}\gamma^{-1}+{C(k,\gamma)}|\omega|^{-2}\right)v'(0)
\exp [2\omega_it]/2.  $ Using the boundary
conditions~\eqref{u_prime_wall_xiU0y} and~\eqref{eqn_w_w_0}, one can easily
show that $i\omega v'^*(0)Dv'(0) + \text{c.c.} = 0$, and thus using
Eq.~\eqref{membrane_eqn}, one arrives at
\begin{equation}
  \frac{d E'_{\text{\tiny W}}}{d t} = \frac{e^{2\omega_it}}{2Re}\left (-dRe|v'(0)|^2 + 3v'^*(0)Dv'(0) - \frac{v'^*(0)D^3v'(0)}{k^2} + \text{c.c.}\right ).
\end{equation}
In summary, Eq.~\eqref{E_teqn} can be written as the sum of four distinct
terms: $ {d E'}/{d t} = e^{2\omega_it}\sum_{j = 1}^4 \dot{E}'_j, $ where the
$\{\dot{E}'_j\}_{j=1,\cdots,4}$ are the rates of energy transfer via different
routes, namely by mean-shear ($\dot{E}'_1$), viscous dissipation
($\dot{E}'_2$), wall-damping ($\dot{E}'_3$), and by the forcing associated
with the normal stress on the wall ($\dot{E}'_4$). The explicit expressions
for these four terms are
\begin{align}
  \dot{E}'_1 &= -i(2k^2)^{-1}\int_0^\infty DU_{0}\left(\alpha v'^*Dv'+ \beta \eta'^*v'\right)\,dy+\text{c.c.}, \label{Eprimedot_1} \\
  \dot{E}'_2 &= -(2k^2Re)^{-1}\int_0^\infty\left[\eta'^*(k^2-D^2)\eta'+v'^*(k^2-D^2)^2v'\right]dy+\text{c.c.},\\
  \dot{E}'_3 &= -d|v'(0)|^2,\ \dot{E}'_4 =
  (2k^2Re)^{-1}\left\{3k^2v'^*(0)Dv'(0) - {v'^*(0)D^3v'(0)} \right\} +
  \text{c.c.}.  \label{Eprimedot_4}
\end{align}
In Eqs.~\eqref{Eprimedot_1}--\eqref{Eprimedot_4}, the eigenfunctions are
employed after normalizing them such that they have a unit initial energy,
$E'(0)=1$.
\begin{figure}[htbp]
  \begin{center}
    \includegraphics[width=0.9\textwidth]{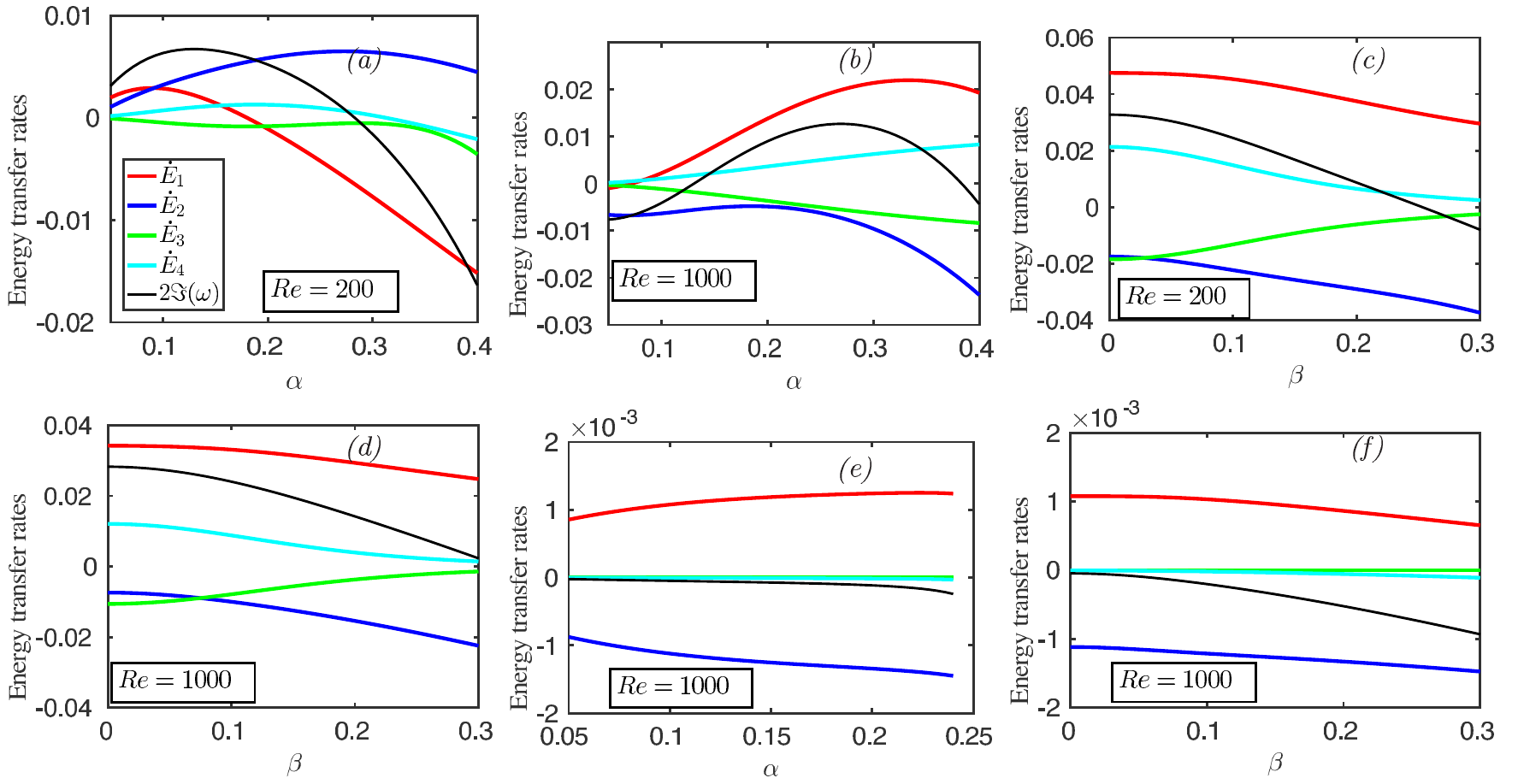}
    \caption{Energy transfer rates in terms of constituent channels: Red ---
      transfer-rate by mean shear; Blue --- transfer rate by viscous
      dissipation; Green --- the transfer rate by wall-damping; Cyan --- the
      transfer rate via forcing by normal stress on the membrane; Black ---
      the total transfer rate: (\textit{a}) ``transitional'' mode with $Re =
      200$, $\beta = 0$, $m=2$, $d=10$, $B = 0.2$, $T = 0.1$, $K = 0.05$;
      (\textit{b}) TS mode with parameters same as (\textit{a}), but for $Re =
      1000$; (\textit{c}) transitional mode with $Re = 200$, $\alpha = 0.25$,
      $m=2$, $d=1$, $B = 0.2$, $T = 0.1$, $K = 0.05$; (\textit{d}) Same as
      (\textit{c}) but for $Re = 1000$; (\textit{e}) TWF mode with $Re =
      1000$, $\beta = 0$, $m=2$, $d=0$, $B = 0.2$, $T = 0.1$, $K = 0.05$;
      (\textit{f}) TWF with $Re = 1000$, $\alpha = 0.1$, $m=2$, $d=0$, $B =
      0.2$, $T = 0.1$, $K = 0.05$.}
    \label{fig:eratebudget}
  \end{center}
\end{figure}

%
Figure~\ref{fig:eratebudget} shows the variations of each of these rates of
energy transfer $\dot{E}'_j$ with respect to $\alpha$ and $\beta$. To analyze
the two different modes of instabilities (i.e., due to transitional modes and
TS modes), where one is present at very low Reynolds numbers and the other in
the regime of low to large $Re$, we carry out an energy budget analysis for
two values of $Re$, specifically $Re = 200$ and $Re = 1000$. When the analysis
is made with respect to $\alpha$, we fix $\beta = 0$, and when it is made with
respect to $\beta$, we fix $\alpha = 0.25$.  Also shown in
Fig.~\ref{fig:eratebudget} is the total energy-transfer-rate, $\dot{E}' \equiv
2\Im(\omega)$.

Figure~\ref{fig:eratebudget}(\textit{a}) shows the analysis of the
transitional mode with respect to $\alpha$ for $Re = 200$. This corresponds to
the analysis along the line $Re = 200$ in
Fig.~\ref{fig:ci_rexal_comp}(\textit{b}). It reveals that in the region of
$\alpha$ considered, viscous effects have a clear destabilizing effect on the
flow, as attested by the positive values of the energy transfer rate
$\dot{E}'_2$. However, it is worth noting that this rate becomes negative in
the same range of $\alpha$ but at higher $Re$ as shown in
Fig.~\ref{fig:eratebudget}(\textit{b}) where an analysis of a TS mode is
carried out. This brings to light the dual nature of viscous effects for this
particular flow at various Reynolds numbers. At low $Re$ (see
Fig.~\ref{fig:eratebudget}(\textit{a})), the rate of energy transfer from the
meanflow ($\dot{E}'_1$) is such that it contributes to the instability at very
low values of $\alpha$, and \textit{vice versa} at high values of
$\alpha$. Therefore, at low $Re$, two-dimensional perturbations are prone to
long wave instability caused by the mean-shear. However, as evident from
Fig.~\ref{fig:eratebudget}(\textit{b}), the mean-shear predominantly causes an
instability for a wider band of wavenumbers at high $Re$ where the
lease-decaying mode is of TS kind.

The transfer rates due to the compliant wall, $\dot{E}'_3$ (wall-damping) and
$\dot{E}'_4$ (fluid-forcing), cancel each other at low $Re$ and low $\alpha$
as shown in Fig.~\ref{fig:eratebudget}(\textit{a}). However, in the regime of
low $Re$ and high $\alpha$, the compliant wall has a clear stabilizing effect
as attested by the negative values of both $\dot{E}'_3$ and $\dot{E}'_4$. In
the high-$Re$ regime (see Fig.~\ref{fig:eratebudget}(\textit{b})), both of
these terms are almost canceling each other for all values of $\alpha$.

The energy transfer rate due to viscosity, $\dot{E}'_2$ at low $Re$ turns
negative upon reducing the wall-damping parameter $d$, as evident from the
comparison of Fig.~\ref{fig:eratebudget}(\textit{a}) ($d = 10$) and
Fig.~\ref{fig:eratebudget}(\textit{c}) ($d = 1$). From these subfigures, one
can notice that the signs are opposite at the common point of $(\alpha, \beta)
= (0.25, 0)$.

Upon increasing the obliquity (i.e., the spanwise wavenumber $\beta$) of the
perturbations, the qualitative nature of the variations of these individual
terms remains constant for low, as well as high $Re$, as displayed in
Figs.~\ref{fig:eratebudget}(\textit{c}) \&~\ref{fig:eratebudget}(\textit{d}).

Figures~\ref{fig:eratebudget}(\textit{e}) \&~(\textit{f}) show these four
energy transfer rates for a TWF mode propagating downstream. As the TWFs have
significant growth/decay rate at low $d$, we have set $d = 0$, and have shown
these rates as a function of $\alpha$ in
Fig.~\ref{fig:eratebudget}(\textit{e}) and $\beta$ in
Fig.~\ref{fig:eratebudget}(\textit{f}). One can note that this TWF-d mode is
always stable given the negative net-rate ($2\Im\{\omega\}$) for all $\alpha$
and $\beta$. This is in good agreement with its classification as a Class-B
mode~\cite{carpenter1990status,gad1996compliant}, which allows it to be either
negative-energy or positive-energy waves. In the present case this TWF-d is a
negative-energy wave which causes irretrievable loss of energy at all wave
numbers. Even the component due to the forcing by the fluid on the wall
($\dot{E}'_4$) is negative, which implies that in effect the fluid interaction
actually dampens the wall motion.

For all these different kinds of modes, we found that the sign of the sum of
$\dot{E}'_3$ and $\dot{E}'_4$ alone predicts qualitatively whether the flow is
stable or otherwise at a chosen $\alpha$ or $\beta$ (not shown here, but a
careful look on the green and cyan curves of the panels of
Fig.~\ref{fig:eratebudget} makes this point clear). This phenomenon can be
explained through an analogy with the dynamics of a forced-damped
pendulum. Such an analogy is suggested by the fact that the governing equation
of the wall [see Eq.~\eqref{membrane_eqn}] is a simple harmonic equation with
both forcing and damping terms.

Since the forcing term $\sigma'_{yy}$ is complex, its phase can be tuned by
the variations in $Re$, $\alpha$ and $\beta$. This forcing term plays a
crucial role in causing stability/instability, as it can either enhance or
hamper the growth of the wall displacement depending on the value of its
phase. Equation~\eqref{membrane_eqn} can be written as $
\left[-{m\gamma^{-1}\omega^2} + C(k,\gamma)\right] \xi' = f', $ where $f' =
\sigma'_{yy}+i\omega d\xi'$ plays the effective role of a forcing term. Let us
write $f' = |f'| e^{i\phi_1}$ and $v'(0) = |v'(0)| e^{i\phi_2}$, with the
phase difference being $\phi = \phi_1 - \phi_2$. When $\phi = 0$, a resonant
instability occurs as the effective forcing term is in phase with the
velocity. This instability could occur for any $\phi\in (-\pi/2, \pi/2)$ since
the average effective forcing in a cycle yields an increase in $|\xi'|$. The
instability vanishes for $\phi\in (\pi/2, 3\pi/2)$.

\begin{figure}[htbp]
  \centering
  \includegraphics[width=0.7\textwidth]{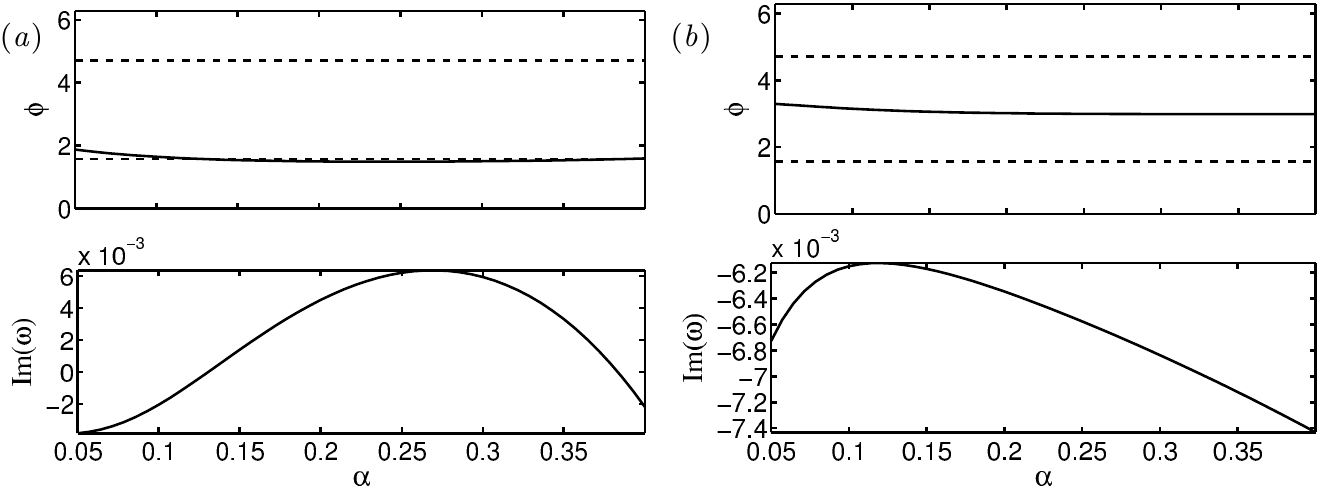}
  \caption{Phase difference, $\phi=(\phi_1-\phi_2)$ (top) and growth rate
    $\Im(\omega)$ (bottom) versus $\alpha$ for $Re = 1000$, $\beta = 0$,
    $m=2$, $d=10$, $B = 0.2$, $T = 0.1$, $K = 0.05$ (same parameters as in
    Fig.~\ref{fig:eratebudget}(\textit{b})): (\textit{a}) least-decaying mode;
    (\textit{b}) second least-decaying mode. The dashed line marks the value
    of $\pi/2$ and $3\pi/2$.}
  \label{phasediffwallvel}
\end{figure}

%
Figure~\ref{phasediffwallvel} shows the growth rate $\Im(\omega)$ and the
phase-difference $\phi$ versus $\alpha$ for the flow and wall parameters set
as identical to those in
Fig.~\ref{fig:eratebudget}(\textit{b}). Figure~\ref{phasediffwallvel}(\textit{a})
shows these results for the least-decaying (i.e., the most unstable) mode and
Fig.~\ref{phasediffwallvel}(\textit{b}) is for the second least-decaying
mode. As expected, when the phase difference is such that $|\phi|<\pi/2$, the
growth rate turns positive for a wide range of values of $\alpha$ as is shown
in Fig.~\ref{phasediffwallvel}(\textit{a}). At both ends of the instability
regime, i.e., as the growth rate approaches zero in
Fig.~\ref{fig:eratebudget}(\textit{a}), the phase difference approaches the
value of $\pi/2$. When $|\phi| = \pi/2$, the effective forcing term $f'$
produces an increase in $\xi'$ for half of a cycle, and equally hinders it for
the other half, thereby causing marginal stability.

The situation is reversed for the second least-decaying mode shown in
Fig.~\ref{phasediffwallvel}(\textit{b}). In the $\alpha$-range of stability,
as anticipated, we obtain $|\phi|<\pi/2$. Here, the stability occurs as the
forcing due to the normal stress is opposing the oscillation of the wall for
the most part of the cycle. We confirm that this result holds for the other
modes as well.

\section{Transient growth study}
\label{sec:transient}

Since the linearized Navier--Stokes operator is nonnormal, the eigenfunctions
can be nonorthogonal, thus resulting in a transient temporal growth of the
norm of a superposition of such eigenfunctions. However, it is worth noting
that each term of such a superposition, taken individually, can be
asymptotically stable. Such a transient growth is inviscid for flows over
solid surfaces, where the viscosity hinders the growth only at a later
time. It has been established that in the case of the boundary layer flow over
a solid wall under parallel flow approximation, the temporal maximum of the
transient growth is proportional to $Re^2$ for streamwise independent modes
(i.e., for $\alpha = 0$)~\cite{gustavsson1991energy}. Such transients are
responsible for the transition to turbulence in the subcritical regime by
amplifying the infinitesimal perturbation to sufficiently finite amplitude
susceptible to the action of nonlinearity.

Let us consider the following superposition of modes
\begin{equation}
  \mathbf{\tilde{q}} = \sum_{n=1}^{K_1}\kappa_{n} \mathbf{q'}_n(y; \alpha, \beta) \exp (i\omega_n t), \ \mbox{with} \ \mathbf{q'}_n = \{v'_n, \eta'_n, v'_n/\omega_n \}^{\rm T},
  \label{superpose_def}
\end{equation}
where the subscript $n$ runs over a selected set of $K_1$ eigenfunctions, and
$\left\{\kappa_n\right\}_{n=1,\cdots,K_1}$ are constants. The integer $K_1$ is
chosen such that~\eqref{superpose_def} includes a sufficient number of modes
relevant to the transient growth. This latter point is further discussed in
what follows. To quantify the size of $\mathbf{\tilde{q}}$, we use the
definition~\eqref{E_def} for the total energy of the system (fluid$+$wall),
which is recast as $ \tilde{E}(t) =
(2k^2)^{-1}\int_{0}^{\infty}\mathbf{\tilde{q}}^{\transp}\mathbf{M}\mathbf{\tilde{q}}\,
dy, $ where the matrix $\mathbf{M}$ is
\begin{equation}
  \mathbf{M} = 
  \left ( 
    \begin{array}{ccc}
      k^2 + D^{\rm T}D + \delta(y)k^2m/\gamma & 0 & 0\\
      0 & 1 & 0 \\
      0 & 0 & \delta(y)k^2C(k, \gamma)
    \end{array}
  \right ).
  \label{E_def_1}
\end{equation}
Following the same methodology as in~\cite{schmid2001stability}, the growth
rate at an instant $t$ is defined as $ G(t) \equiv
\max_{\kappa_n}({\tilde{E}(t)}/{\tilde{E}(0)})=\|\mathbf{F} \exp(i{\boldsymbol
  \Omega} t)\mathbf{F}^{-1}\|_2, $ where ${\boldsymbol \Omega} =
\mbox{diag}\{\omega_n\}$ where ${n=1,\cdots,K_1}$,
$\mathbf{F}^{\transp}\mathbf{F} = \mathbf{A}$ with $A_{ij}
=(2k^2)^{-1}\int_{0}^{\infty}\mathbf{q'}_j^{\transp}\mathbf{M}\mathbf{q'}_i\,
dy$.  Furthermore, the maximum growth rate is given by $G_{\max} \equiv
\max_{t>0} G(t)$ with $t_{\max}$ defined such that $G(t_{\max}) = G_{\max}$.

We now present results on transient growth and optimal perturbations for small
values for $\alpha$ and $\beta$ to ward off the pseudospectra.
\begin{figure}[htbp]
  \begin{center}
    \includegraphics[width=0.9\textwidth]{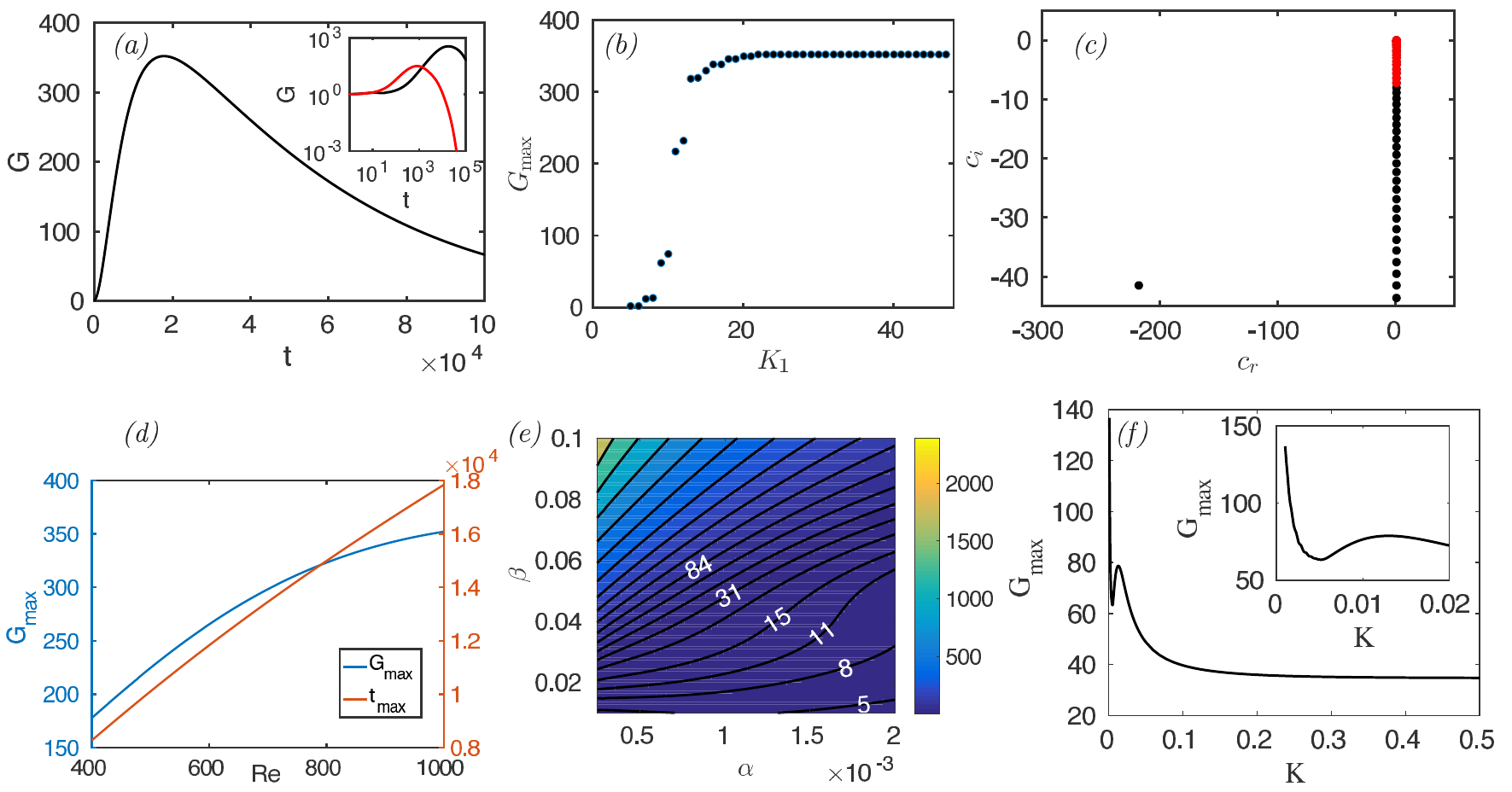}
    \caption{Transient growth: (\textit{a}) Black line is $G(t)$ for $Re =
      1000$, $m = 2$, $d=1$, $K = 0.3$, $B = 3.2$, $T = 0.75$ $\alpha = 0.001$
      and $\beta = 0.09$; the red line in the inset corresponds to the
      solid-wall case; (\textit{b}) $G_{\text{\scriptsize max}}$ versus no. of
      modes $K_1$ in the superposition; (\textit{c}) The red-part of the
      spectrum contributes to 99\% of $G_{\max}$; (\textit{d})
      $G_{\text{\scriptsize max}}$ and $t_{\text{\scriptsize max}}$ versus
      $Re$; (\textit{e}) $G_{\max}$ in $\alpha-\beta$ plane with other
      parameters as in (\textit{a}); and, $G_{\text{\scriptsize max}}$ and
      versus $K$.}
    \label{fig:tg}
  \end{center}
\end{figure}
Figure~\ref{fig:tg}(\textit{a}) shows the growth rate $G(t)$ of modes that are
approximately independent of the streamwise direction (i.e., with $\alpha
\approx 0$). The wall- and flow-parameters have been set in the stable
regime. It should be noted that $t_{\max}$ is nearly $2\times 10^4$, which is
in contrast to that of a solid wall where $t_{\max}\approx
778$~\cite{butler1992three}.  Also shown in the inset of
Fig.~\ref{fig:tg}(\textit{a}) is the comparison with the solid-wall case (red
line).

As $t_{\max} \sim \mathcal{O}(10^4)$ (see Fig.~\ref{fig:tg}(\textit{a})),
which is a fairly large value , the transient growth can be fully described by
a set of least-decaying modes. For this reason, we estimate that considering
$K_1 = 50$ modes is vastly sufficient for all our calculations given that low
values are considered for $\alpha$ and
$\beta$. Figure~\ref{fig:tg}(\textit{b}) shows the contribution of each mode
(in the order of their decay-rate) to $G_{\max}$, thus showing the convergence
with respect to the number of modes.  Figure~\ref{fig:tg}(\textit{c}) shows
the region of the spectrum (marked red) that contributes most to the transient
growth.

Figure~\ref{fig:tg}(\textit{d}) shows the dependence of $G_{\max}$ and
$t_{\max}$ on the Reynolds number. Though $t_{\max}$ varies linearly with $Re$
as in the solid-wall case, the maximum growth rate $G_{\max}$ does not
preserve the quadratic growth with respect to $Re$, which is a trend commonly
observed for parallel flows over solid walls~\cite{gustavsson1991energy}. With
our soft compliant wall, $G_{\max}$ increases at a rate slower than a linear
dependence or $Re$. It is worth noting that the quadratic growth in the
solid-wall case is theoretically supported for such streamwise-independent
modes~\cite{kreiss1994bounds}. However, in the compliant-wall case, the
wall-boundary condition has a significantly more complex dependence on the
Reynolds number. This results in an inability to absorb the entire system's
$Re$-dependence by a scaling of the dependent variables---a step required in
order to establish the quadratic dependence on $Re$.

Figure~\ref{fig:tg}(\textit{e}) shows the contours of $G_{\max}$ in the
$\alpha-\beta$ plane. The supremum of $G_{\max}$ appears for a mode with
vanishing $\alpha$. This fact is consistent with what is observed with the
boundary layer flow over a solid wall, and one can expect streaks to form the
optimal patterns. A contrasting aspect is the rapid increase in $G_{\max}$
with respect to $\beta$, which is at much higher rate than that of such flow
over a solid wall~\cite{schmid2000linear}. Such rapid increase is a signature
of the higher nonnormality of the underlying linear operator than that of the
flat-rigid wall case. This causes the computation to become more sensitive to
the accuracy of the pseudospectra even at such low values of wavenumbers. As a
consequence, with the current double-precision accuracy, it is impossible to
capture the optimal value for $\beta$.

Figure~\ref{fig:tg}(\textit{f}) displays the variation of $G_{\max}$ with
respect to the stiffness constant $K$ for three-dimensional modes. Overall,
$G_{\max}$ decreases when the wall undergoes transition from a soft one (i.e.,
with low $K$) to a hard one, and this trend saturates for $K>0.3$. This
suggests that, in the case of a soft compliant wall, there exists a mechanism
of transient growth apart from the inviscid growth that normally exists in the
solid-wall case. We anticipate this to be a combination of pressure and
viscous forces on the wall. As the transient growth is predominantly due to
the nonorthogonality of continuous modes within the boundary layer, the
difference in the $G_{\text{\scriptsize max}}$ between the solid- and
compliant-wall cases should hail from these modes. These continuous modes have
a nonvanishing boundary behavior as can be see in
Fig.~\ref{fig:spectra}(\textit{e}). Therefore, the inner product of these
functions can naturally be higher compared to the solid-wall case, where these
functions have a vanishing boundary behavior. Thus, the wall compliance
enhances the nonorthogonality of the eigenfunctions, making them susceptible
to a stronger transient growth. The role of viscosity in enhancing the
transient growth is further confirmed in Sec.~\ref{sec:transient-energy} by
means of a term-by-term energy budget analysis.

\subsection{Optimal patterns and their origins}

\begin{figure}[htbp]
  \begin{center}
    \includegraphics[width=0.8\textwidth]{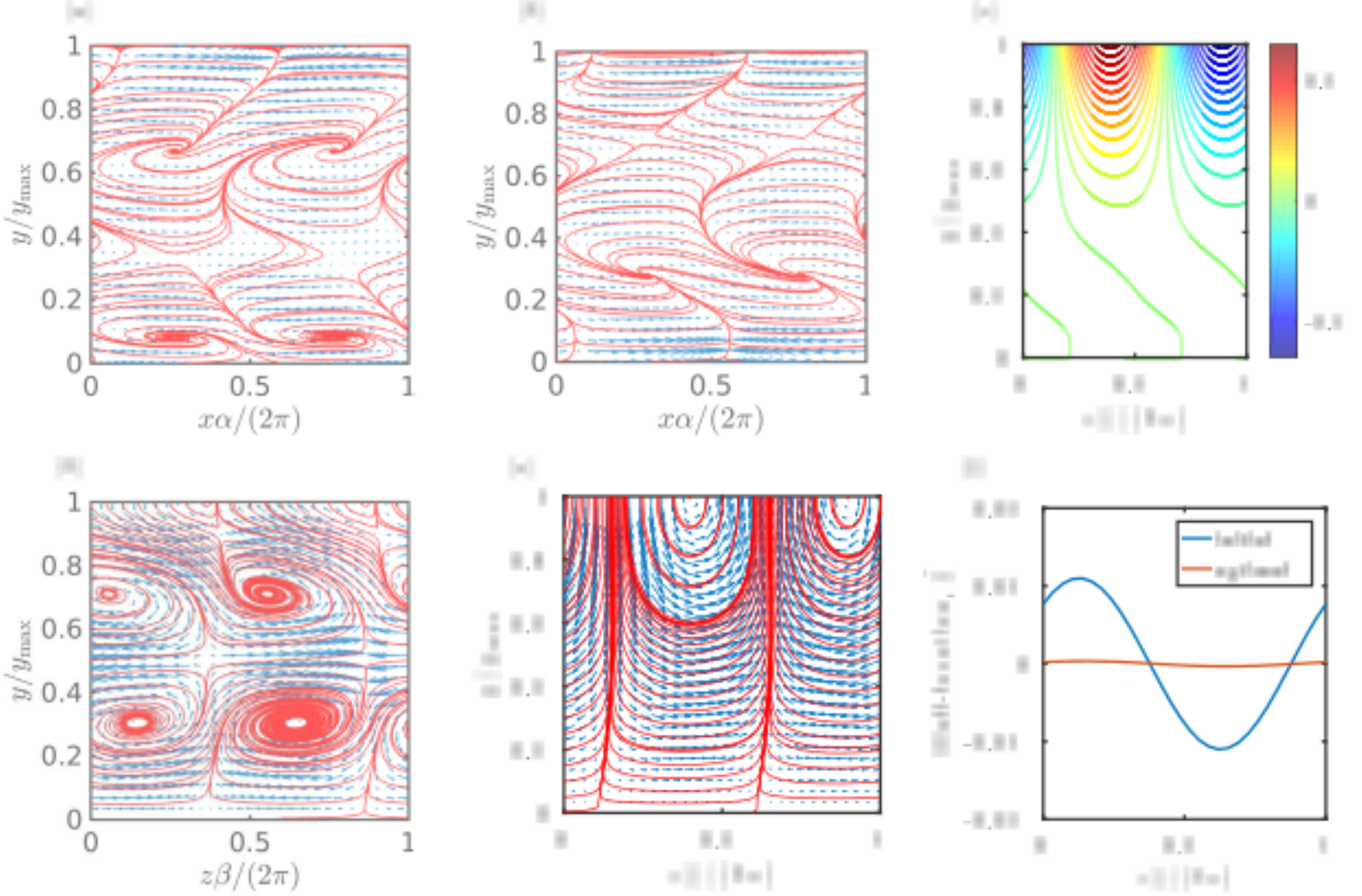}
    \caption{Initial and optimal perturbations for $Re=1000$, $m=2$, $d=1$,
      $K=0.3$, $B = 3.2$, $T=0.75$: (\textit{a}) and (\textit{b}) are initial
      and optimal patterns of perturbation velocities, respectively, with
      $\alpha=0.002$, $\beta = 0$ in $(x,y)$ plane (Red lines show
      streamlines); (\textit{c}) is contours of streamwise optimal
      perturbation velocity in $(y,z)$ plane with $\alpha=0.00025$, $\beta =
      0.09$; (\textit{d}) and (\textit{e}) initial and optimal patterns of
      perturbation velocities, respectively, corresponding to (\textit{c});
      (\textit{f}) the initial and optimal wall fluctuation.}
    \label{fig:opt}
  \end{center}
\end{figure}

%
We now turn to the problem of finding the coefficients
$\{\kappa_n\}_{n=1,\cdots,K_1}$ in Eq.~\eqref{superpose_def} that take the
superposition $\mathbf{\tilde{q}}$ to the state of maximum energy, i.e.,
$\tilde{E}(t_{\max}) = G_{\max}$. This set of $\{\kappa_n\}_{n=1,\cdots,K_1}$
satisfying $E(t_{\max}) = G_{\max}$ can be found by means of
$\greekvektor{\kappa} = \mathbf{F}^{-1}\mathbf{U}(:,1)$, where $\mathbf{U}$ is
the unitary matrix arising from the following singular value decomposition:
$\mathbf{F} \exp(i{\boldsymbol \Omega} t) \mathbf{F}^{-1} =
\mathbf{USV^{\transp}}$. With the obtained vector ${\boldsymbol
  \kappa}=(\kappa_1,\cdots,\kappa_{K_1})^{\transp}$, we have
$\mathbf{\tilde{q}}(t =0)$ denoted as the \textit{initial perturbation} and
$\mathbf{\tilde{q}}(t = t_{\max})$ as the \textit{optimal
  perturbation}. Figures~\ref{fig:opt}(a) \&~\ref{fig:opt}(b) show the initial
and optimal patterns of perturbation velocities in the $(x,y)$ plane, while
Fig.~\ref{fig:opt}(d) \&~\ref{fig:opt}(e) show the same in the $(y,z)$ plane.

\begin{figure}[htbp]
  \begin{center}
    \includegraphics[width=0.5\textwidth]{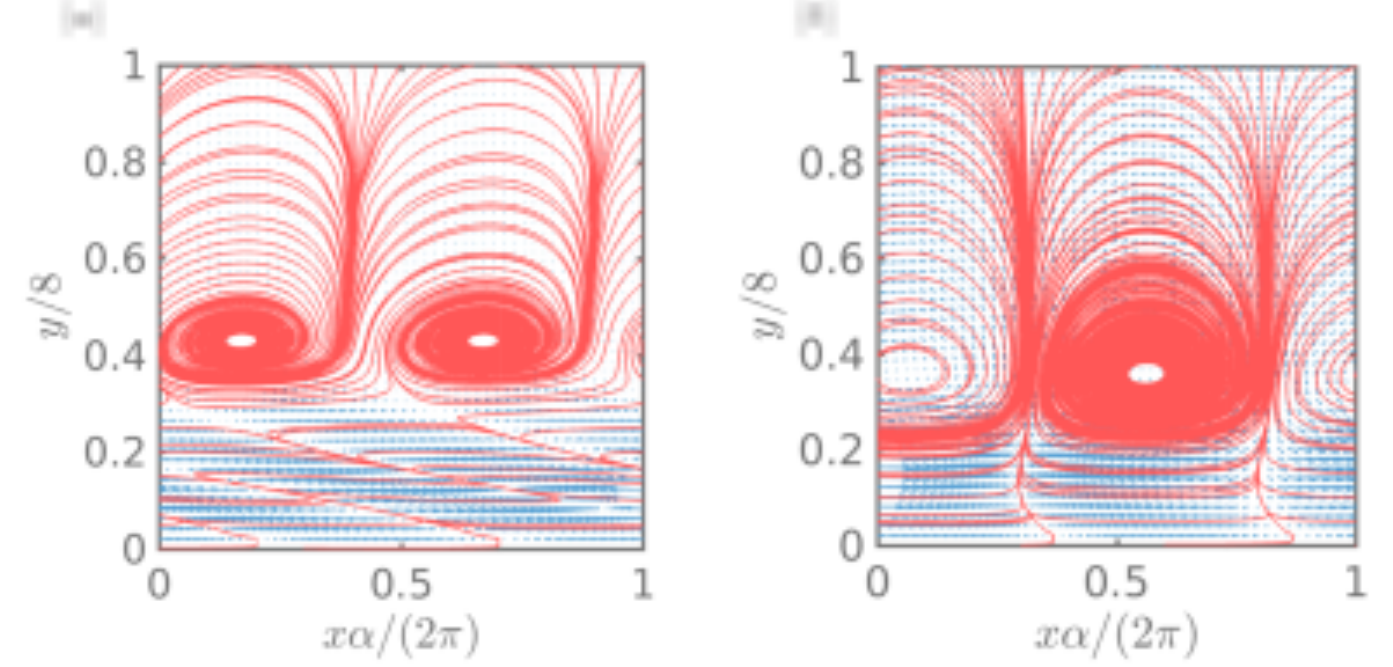}
    \caption{Initial (\textit{a}) and optimal (\textit{b}) perturbations in
      the case of flat rigid wall with $Re=1000$, $\alpha=0.1$, and $\beta =
      0$ in $(x,y)$ plane (Red lines show streamlines).}
    \label{fig:opt_solid}
  \end{center}
\end{figure}

In the case of 2D modes, the initial and optimal patterns are expected to
follow the classical Orr-mechanism. To contrast against the flat-rigid wall
case, we show the optimal patterns for such a case in Fig.~\ref{fig:opt_solid}
with $\alpha = 0.1$. We have chosen such large value of $\alpha$, as the
flat-rigid wall boundary layer does not exhibit transient growth at the value
of $\alpha$ of Fig.~\ref{fig:opt}(a). In the flat-rigid wall case, the initial
pattern associated with the 2D modes is such that the perturbation velocities
in the $(x,y)$ plane are in the direction opposing the mean-shear, as shown in
Fig.~\ref{fig:opt_solid}(a). The streamlines are tilted against the meanflow
within the boundary layer. This perturbation structure gets tilted in the
direction of the mean-shear at the later optimal time. However, this
particular phenomenon is absent in the present boundary layer flow over a
compliant wall as shown in Fig.~\ref{fig:opt}(a) \&~\ref{fig:opt}(b). This is
because the circulation is not conserved in the present case, even in the
inviscid limit. Such a conservation of circulation (i.e., $D\Gamma/Dt = 0$) is
a prerequisite for the Orr-mechanism to take place: when such conservation is
ensured, the elongated contours tilted against the flow and outlining the
constant vorticity/streamlines in the fluid-material undergoes modifications
by the natural fluid motion such that their lengths get shortened until a
point of time where the maximum of the transient growth happens, thus causing
an increase in the velocity field so as to conserve the circulation.  In the
compliant-wall case, there is a dynamic entry of vorticity into the flow
domain. One can view this wall-dynamics as some sort of forcing on the
velocity of a flow over an otherwise flat-rigid wall, and this force, owing to
time-dependency and wall dissipation, is not conservative. It can be deduced
that the circulation is not conserved in such situation by the same analysis
used to arrive Kelvin's theorem.  Not surprisingly, this results in notably
different initial and optimal patterns.


%
In the case of nearly streamwise-independent modes ($\alpha \approx 0$), the
initial and optimal patterns shown in Fig.~\ref{fig:opt}(d)
\&~\ref{fig:opt}(e) contrast with the observations for the solid-wall
case~\cite{butler1992three} with the cores of the counter-rotating vortices
being located at a farther distance away from the wall. The corresponding
streaks are shown in Fig.~\ref{fig:opt}(c). Figure~\ref{fig:opt}(f) shows the
optimal pattern of the wall in the spanwise direction. The minimum and maximum
of the wall pattern at initial time match with the locations of suction and
impingement of the flow-fields, respectively, at optimal time as shown in
Figs.~\ref{fig:opt}(e) \&~\ref{fig:opt}(f). As time evolves towards the
optimal instant, these suction and impingement decrease the spatial amplitude
of the wall in order to retrieve some potential energy so that the velocity
field can have access to it. From Eq.~\eqref{kinematic_1}, such reduction in
wall amplitude implies a reduction in the streamwise component of the velocity
field. In turn, this causes the streaks to shift upwards as observed.  The
structure at optimal time is similar to that of the rigid-wall case---except
for the fact that is farther away from the wall---due to the lift-up
mechanism: the normal velocity induces the normal vorticity. We believe that
the observed phenomena is an interactive effect with the wall. The fluid
energy near the wall is minimum as one can see from the lengths of the arrows
in Fig.~\ref{fig:opt}(e) and the contour levels in Fig.~\ref{fig:opt}(c). As
there is little variations in the magnitude of vectors and contour levels, one
can also see that the viscous dissipation is very low for such optimal
patterns.

\subsection{Transient energy budget analysis}
\label{sec:transient-energy}

In the same spirit as the previous modal study, we consider carrying out a
transient energy budget study for the superposition of modes
$\mathbf{\tilde{q}}$, as this will help us gain insight into the mechanisms of
the transient growth. Here, the method described in~\cite{malik2008linear} is
followed to calculate nonmodal energy. When compared with the solid-wall case,
the transient growth characteristics for the compliant-wall case has primarily
two contrasting features: (\textit{i}) the transient amplification is larger
though slower compared to the solid-wall case (see
Fig.~\ref{fig:tg}(\textit{a})); (\textit{ii}) for modes with vanishing
$\alpha$, apart from the collapse of the scaling $G_{\max}\sim Re^2$, the
growth rate of $G_{\max}$ with respect to $Re$ decreases with increasing $Re$
as can be seen from Fig.~\ref{fig:tg}(\textit{d}) (i.e. $d^2G_{\max}/dRe^2 <
0$). We explain these behaviors in the following together with other
observations.

The present state of superposition of modes $\mathbf{\tilde{q}}$ in
Eqs.~\eqref{Eprimedot_1}--\eqref{Eprimedot_4} can be rewritten as
\begin{eqnarray}
  \tilde{E}_1(t) &=& \tilde{E}_1(0)+(2k^2)^{-1}\sum_{j,k} \nolimits S_{jk} \int_0^\infty DU_{0}\left(\alpha v'^*_kDv'_j+ \beta \eta'^*_kv'_j\right)\,dy+\text{c.c.}, \label{Etilde_1} \\ 
  \tilde{E}_2(t) &=& \tilde{E}_2(0)-i(2Re k^2)^{-1}\sum_{j,k} \nolimits S_{jk}\int_0^\infty \left[\eta'^*_k(k^2-D^2)\eta'_j+v'^*_k(k^2-D^2)^2v'_k\right]+\text{c.c.},\\ 
  \tilde{E}_3(t) &=& \tilde{E}_3(0)-id\sum_{j,k} \nolimits S_{jk}v'^*_k(0)v'_j(0),\\ 
  \tilde{E}_4(t) &=& \tilde{E}_4(0)+i(2Re)^{-1}\sum_{j,k} \nolimits S_{jk}\left\{3v'^*_k(0)Dv'_j(0) -k^{-2} v'^*_k(0)D^3v'_j(0)\right\} + \text{c.c.}, \label{Etilde_4}
\end{eqnarray}
where $ S_{jk} = {\kappa_k^* \kappa_j}\left\{\exp\left[-i(\omega_j -
    \omega^*_k)t\right]-1\right\}/{(\omega_j - \omega^*_k)}, $ $\tilde{E}_1$
is the energy from the meanflow, $\tilde{E}_2$ is the energy loss due to
viscous dissipation, $\tilde{E}_3$ is the energy dissipated by wall-damping
and $\tilde{E}_4$ is the energy transferred to the wall by means of the fluid
interaction. The initial values $\{\tilde{E}_j(0)\}_{j=1,\cdots,4}$ can be
chosen such that the total energy is equal to the unity. Without any loss of
generality, we choose $\tilde{E}_j(0) = \{1,0,0,0\}$, as these are additive
constants and that the temporal evolution far greater than the initial values.

\begin{figure}[htbp]
  \centering
  \includegraphics[width=0.9\textwidth]{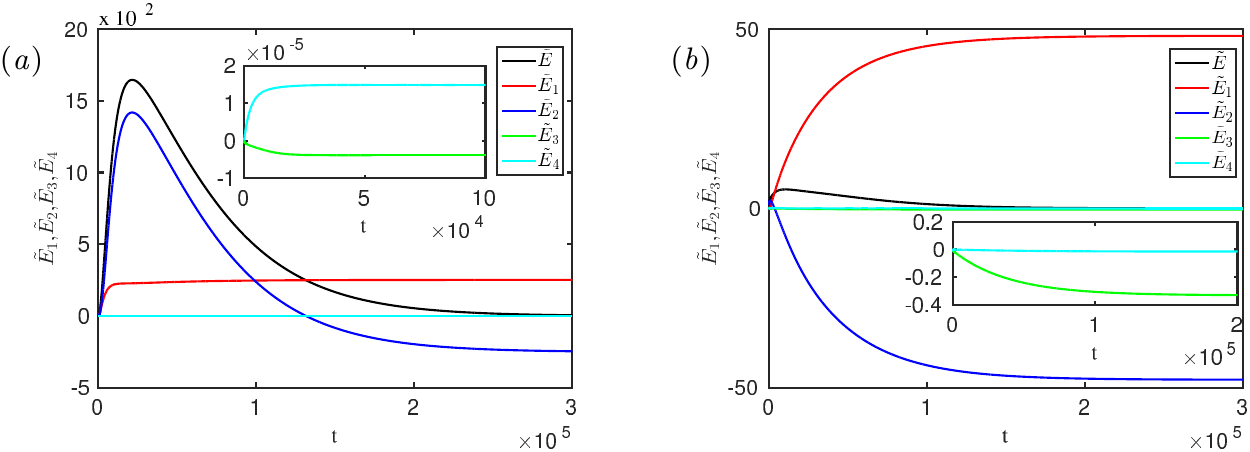}
  \caption{Transient energy budget for $Re=1000$, $m=2$, $d=1$, $K=0.3$, $B =
    3.2$, $T=0.75$: (\textit{a}) $\alpha=2.5\times10^{-4}$, $\beta = 0.09$;
    (\textit{b}) $\alpha=0.002$, $\beta = 0.01$.}
  \label{fig:ebudget_nm}
\end{figure}

Figure~\ref{fig:ebudget_nm} shows these components of the total energy for two
sets of $(\alpha, \beta)$, which are chosen such that they are points in very
distinct areas of $G_{\max}=f(\alpha,\beta)$ in
Fig.~\ref{fig:tg}(\textit{e}). First, let us consider the case of vanishing
$\alpha$ as in Fig.~\ref{fig:ebudget_nm}(\textit{a}). Surprisingly, and unlike
the solid-wall case, the viscous dissipation term ($\tilde{E}_2$) enhances the
transient growth in a manner dominant than the mean-shear ($\tilde{E}_1$). The
reason for this finds its origin in the behavior of the eigenfunctions of the
continuous spectrum within the boundary layer (see
Fig.~\ref{fig:spectra}(\textit{e})). Specifically, in the region $0\leq y<5$,
some of these functions---besides having a nonperiodic behavior with respect
to $y$ unlike in the freestream---have larger values for higher-order
derivatives in comparison to the solid-wall case. In the solid-wall case,
though the freestream oscillations are quenched by the mean-shear
(shear-sheltering), the values of the higher-order derivatives are small. In
the present compliant-wall case, the enhancement in these values due to the
specific wall-dynamics results in an increased nonorthogonality. Indeed, the
inner-product of the modes with the eigenfunctions in this region could be
higher as well. In turn, this contributes to the transient growth. However, as
shown in Fig.~\ref{fig:ebudget_nm}(\textit{a}), these terms eventually
dissipate the energy in the superposition state asymptotically.

However, it is worth noting that the nonsymmetric appearance of $\alpha$ and
$\beta$ in Eqs.~\eqref{dynamic} \&~\eqref{u_prime_wall_xiU0y} leads to
intricate viscosity-dependent effects. Indeed, the actual effects of viscosity
depend on the values of $(\alpha, \beta)$ and how they affect the distribution
of the terms $D^2v'$, $D^4v'$, $D^2\eta'$. In the situation where $(\alpha,
\beta)$ lie closer to the line of $\beta = 0$ in
Fig.~\ref{fig:tg}(\textit{e}), the predominant transient growth is due to the
energy transfer from the meanflow to the superposition-state (see
Fig.~\ref{fig:ebudget_nm}(\textit{b})). The reason for this reversed role of
the viscous terms on the transient growth will be further discussed later in
this section.

Figures~\ref{fig:ebudget_nm}(\textit{a}) \& \ref{fig:ebudget_nm}(\textit{b})
show that wall-damping effects monotonously remove energy, while the forcing
by fluid stress on the wall produces both enhancing as well as adverse effects
on the transient growth depending on the values of $(\alpha, \beta)$. These
results can be understood from the exact solution of the wall equation
\begin{eqnarray}
  \tilde{\xi}(t) &=& \frac{F_2 -\lambda_2 F_1}{\lambda_1 - \lambda_2} e^{\lambda_1 t} + \frac{\lambda_1 F_1 -F_2}{\lambda_1 - \lambda_2} e^{\lambda_2 t} + \sum_j \frac{\kappa_j\sigma'_{22,j}}{(\lambda_1 + i\omega_j)(\lambda_2 + i\omega_j)} e^{-i\omega_j t}, \label{shm_soln}\\
  \mbox{where} \ \ F_1 &=&\sum_j\kappa_j\left [ \frac{iv'_j(0)}{\omega_j} - \frac{\sigma'_{22,j}}{(\lambda_1 + i\omega_j)(\lambda_2 + i\omega_j)} \right ], \ 
  F_2 =\sum_j\kappa_j\left [ v'_j(0) + i\omega_j \frac{\sigma'_{22,j}}{(\lambda_1 + i\omega_j)(\lambda_2 + i\omega_j)} \right ] \nonumber,\\
  \mbox{and} \ \ \lambda_{1,2} &=& \frac{1}{2m}\left[-\gamma d\pm\sqrt{(\gamma d)^2-4m\gamma
      C(k,\gamma)}\right] . \label{wallgrowthrate}
\end{eqnarray}
Note that this solution satisfies the initial conditions $\tilde{\xi}(0) =
i\sum_j\kappa_j v'_j(0)/\omega_j$ and $\tilde{\xi}_t(0) =
\sum_j\kappa_jv'_j(0)$.

In this solution~\eqref{shm_soln}, the homogeneous parts (i.e., the first two
terms) decay due to the effect of the damping coefficient $d$. However, the
imaginary part of the two exponents $\lambda_1$ and $\lambda_2$ differ only by
a sign, thereby implying that these two homogeneous parts are orthogonal. This
explains the absence of transient growth in the energy lost by the
wall-damping in Fig.~\ref{fig:ebudget_nm}. Furthermore, the
solution~\eqref{shm_soln} exhibits transient growth only from the particular
solution, which is the response to the forcing by normal stress. As this part
of the solution involves summing over the eigenmodes---in which each of them
have their own modal temporal evolution---it collectively exhibits transient
growth. This explains the presence of transient growth of the energy in
Fig.~\ref{fig:ebudget_nm}(a) due to the term representing forcing by normal
stress. Therefore, the root of the transient growth in the displacement of the
wall hails from the flow.

Another interesting observation is the fact that when the forcing by the fluid
enhances the transient growth of the wall-displacement, so do the viscous
effects on the transient growth in flow. Let us recall here that the
displacement $\tilde{\xi}$ and velocity
$(\tilde{u},\tilde{v},\tilde{w})^{\transp}$ are related through the dynamic
and kinematic boundary conditions (see Sec.~\ref{sec:boundary}), and the
continuity equation. A transient growth in $\tilde{\xi}$ due to fluid forcing
as per Eq.~\eqref{shm_soln} will result in a transient growth in the velocity
field in the layer immediately adjacent to the wall. In turn, this transient
growth in that layer of fluid will eventually be transferred to the entire
flow domain via viscosity. This results in viscous terms contributing to the
growth of the overall disturbance of the combined fluid-wall system. However,
when considering the situation where the compliant wall reduces the energy of
the system during transients as in the case of
Fig.~\ref{fig:ebudget_nm}(\textit{b}), the layer of the fluid also loses
momentum in order to satisfy the boundary conditions at the wall. In turn,
this reduction in the magnitude of the velocity field is spread by viscosity
to the entire domain, thereby highlighting the expected and usual dissipative
role of viscous terms for such growth (see
Fig.~\ref{fig:ebudget_nm}(\textit{a})).

These explanations can also be understood with the help of
Eqs.~\eqref{Etilde_1}--\eqref{Etilde_4}. In these equations, as a matter of
convention, let us consider that the dependent variables with
complex-conjugate receives the energy from each term where they appear. In the
case of $\alpha \approx 0$, we can safely assume that the energy is initially
in $v'(y)$ and $w'(y)$ of fluid, as the streaks has not yet formed. The
forcing term,
\begin{equation*}
  T_1 = i(2Re)^{-1}\sum_{j,k} \nolimits S_{jk}\left[3v'^*_k(0)Dv'_j(0) -k^{-2}
    v'^*_k(0)D^3v'_j(0)\right] + \text{c.c.}
\end{equation*}
transfers energy into $v'$, which is being lost in part by the following
damping term
\begin{equation*}
  T_2 = -id\sum_{j,k} \nolimits S_{jk}v'^*_k(0)v'_j(0)+\text{c.c.}.
\end{equation*}
As the wall is being driven, the energy transferred into $v'(0)$ by the
forcing can be transferred to $v'(y)$ for $y>0$ by the following part of the
viscous dissipation term
\begin{equation*}
  T_3 =
  -i(2Re k^2)^{-1}\sum_{j,k}\nolimits S_{jk}\int_0^\infty\left[v'^*_k(k^2-D^2)^2v'_k\right]+\text{c.c.},
\end{equation*}
due to the presence of derivatives. Meanwhile, the mean-shear would transfer
the energy from $v'$ into $\eta'$ through the following term $ T_4 =
(2k^2)^{-1}\sum_{j,k}S_{jk} \int_0^\infty DU_{0} \beta \eta'^*_kv'_j\,
dy+\text{c.c.}, $ especially near the wall, since $DU_{0}$ is maximum at the
wall. This transferred energy into $\eta'(y\simeq 0)$ near the wall, gets
redistributed to $\eta'(y)$ for $y$ located off the wall by means of the part
of the viscous term $ T_5 = -i(2Rek^2)^{-1}\sum_{j,k}S_{jk}\int_0^\infty
\eta'^*_k(k^2-D^2)\eta'_j+\text{c.c.}, $ which is made possible by the
presence of the derivatives in the kernel. This detailed picture reveals the
dual role played by the viscous terms in the mechanisms of transient
growth. In the case of $\beta \sim 0$, the initial and optimal energy is
confined to $u'$ and $v'$. In such situation, the mean-shear transfers the
energy from $v'(y\simeq 0)$ close to the wall into $u'$ of the fluid through
the term $ T_6 = (2k^2)^{-1}\sum_{j,k}S_{jk} \int_0^\infty DU_{0}\alpha
v'^*_kDv'_j\, dy+\text{c.c.}, $ which is immediately being dissipated by a
part of $T_3$, namely $ T_7 \equiv i(Re
k^2)^{-1}\sum_{j,k}S_{jk}\int_0^\infty\left[v'^*_kD^2v'_k\right]+\text{c.c.}
$

The observation in Fig.~\ref{fig:tg}(\textit{d}) that the maximum of the
transient growth $G_{\max}$ is slower than a linear rate with respect to $Re$
can now be explained based on the above discussion. It is apparent that the
viscous terms play a crucial role in the compliant-wall case, in terms of
propagating the transient growth from the wall displacement into the flow
domain. Therefore, the slower rate of increase for $G_{\max}$ with respect to
$Re$ is due to the reduced forcing stress on the compliant wall at high
Reynolds number, as is evident from Eq.~\eqref{sigmaprime}. In the inviscid
limit, the transient growth is entirely due to the flow field without any
enhancement from the wall dynamics, i.e. the fluid-solid interaction occurs
only via one-way coupling as noted before.

Finally, from the solution given by Eq.~\eqref{shm_soln}, one can also
identify the reason behind the fact that the transient growth is weaker in the
solid-wall case in comparison to the compliant-wall one. Indeed, in the limit
of $C(k,\gamma) \rightarrow \infty$, the membrane approaches the conditions of
a solid wall. In this limit, the norm of the denominator $(\lambda_1 +
i\omega_j)(\lambda_2 + i\omega_j)$ in one of the terms in Eq.~\eqref{shm_soln}
tends to blow up, thus weakening the response to forcing. Therefore, in the
solid-wall limit, the major contribution of the present situation vanishes
altogether, and the transient growth is purely limited to the effects
associated with the inviscid feature of the flow.

\section{Conclusions}
\label{sec:conclusions}

In this paper, the temporal modal and nonmodal growth of three dimensional
perturbations in the boundary layer flow over an infinite compliant flat wall
has been investigated. The two key findings from this work are the
following. First, the flow is found to have an instability mechanism
reminiscent of that of a forced and damped harmonic oscillator. We confirmed
that the combined system of fluid and compliant wall have stable/unstable
modes as predicted by the phase of the forcing and damping terms. Second, in
stark contrast to the solid-wall problem, the transient growth in the
compliant-wall case involves a nontrivial contribution of viscous terms, in
particular for the dynamics of streamwise-independent modes. It is found that
this contribution stems from the role of viscous terms in communicating the
transient growth from the wall displacement to the fluid, which in turn stems
from the nonmodal growth of the forcing by the fluid on the wall.

From the formalism standpoint, we approached this problem using a two-variable
formulation---wall-normal velocity and wall-normal vorticity, which
significantly simplifies the treatment of boundary conditions. Specifically,
with this two-variable formulation the quadratic dependence on the eigenvalue
parameter vanishes, thereby drastically reducing the computational effort
required to obtain the spectra. Furthermore, we were able to accurately
compute the discrete and continuous modes from the two separate systems, thus
enabling us to properly filter the pseudospectra so as to obtain
highly-accurate eigenfunctions. Still with the help of this two-variable
formulation, we determined an instance of each of the hydroelastic
modes---static wave divergence, traveling wave flutter and transitional modes,
and analyzed the associated eigenfunctions and their stability through an
energy budget analysis. In addition, we identified the instability regions in
the parameters space and also analyzed the growth rate of hydroelastic and
Tollmien--Schlichting modes. For all cases considered, our approach allowed us
to obtain accurate continuous spectra for large negative values of $c_i$. It
is important to highlight that having access to clean continuous spectra
without the pseudospectra is an essential prerequisite to performing the
transient growth study.

This boundary layer flow over an infinite compliant wall exhibits a higher
transient growth, though at a slower rate, as compared to the same flow over a
solid wall. In the solid-wall case, the maximum growth rate for streamwise
independent modes scales according to $G_{\max}\sim \mathcal{O}(Re^2)$. Here,
in the compliant-wall case, this scaling law breaks down as a consequence of
the nontrivial dependence of the wall-boundary conditions on the Reynolds
number. The maximum of the transient growth, $G_{\text{\scriptsize max}}$,
increases with $Re$ at a sublinear pace. Using a nonmodal energy budget
analysis, this fact was found to be due to a reduced forcing at higher $Re$,
which in turn reduces the above-mentioned viscous effects on the transient
growth. Other results were found to be similar to those observed in the
solid-wall case. For instance, the superposition of streamwise independent
modes have a stronger transient growth as compared to the modes with $\alpha >
0$.

With soft compliant walls, the transient growth is enhanced since there is a
more intense response of the wall to the forcing of the fluid. The analytical
solution to the equation governing the dynamics of the wall allowed us to
exactly identify the root cause of this effect.

Lastly, the initial and optimal patterns of the 2D modes were found not to
exhibit the phenomenon associated with the Orr-mechanism where velocity
fluctuations appear to counter the effect of the mean-shear initially, and the
rotation of these vortical structures end up being aligned with the mean-shear
at the optimal time. This phenomenon is related to the nonconservation of
circulation because of the time-dependent influx of vorticity generated by the
wall dynamics. Nonetheless, the initial and optimal patterns of the velocity
field are similar to those in the solid-wall case, except that the wall
deformation provides a storage of potential energy at the optimal time.

%
%
%

%
\end{document}